\documentclass[acmsmall,screen]{acmart}

\usepackage{fancyvrb}
\usepackage{xurl}
\usepackage{booktabs}
\usepackage{colortbl}
\usepackage{url}
\usepackage{amsmath,amsfonts}
\usepackage{algorithmic}
\usepackage{graphicx}
\usepackage{textcomp}
\usepackage{xspace}
\usepackage{subcaption}
\usepackage{tabularx}
\usepackage{framed}
\usepackage{chngpage}
\usepackage[many]{tcolorbox}
\usepackage{pgfplots}
\pgfplotsset{compat=newest}
\usepackage[group-separator={,}, group-minimum-digits=4]{siunitx}
\usepackage{balance}
\usepackage{enumitem}
\usepackage{sepnum}
\usepackage{multirow}
\usepackage[htt]{hyphenat}

\usepackage{hyperref}

\usepackage{tikz}

\usepackage{pgf-pie}
\pgfplotsset{compat=1.18}

\newcommand{\rqa}{RQ$_1$}
\newcommand{\rqb}{RQ$_2$}
\newcommand{\rqc}{RQ$_3$}
\newcommand{\rqd}{RQ$_4$}
\newcommand{\rqaa}{How do agent-authored commits differ from human commits?}
\newcommand{\rqbb}{How well can we distinguish agent-authored from human commits after removing authorship signatures?}
\newcommand{\rqcc}{How well does detection generalize to agents unseen during training?}
\newcommand{\rqdd}{How does detection performance change over time?}

\newcommand{\AIDev}{AIDev\xspace}
\newcommand{\CopilotFull}{GitHub Copilot coding agent\xspace}
\newcommand{\CopilotShort}{GitHub Copilot\xspace}
\newcommand{\Copilot}{Copilot\xspace}
\newcommand{\ClaudeCode}{Claude Code\xspace}
\newcommand{\Cursor}{Cursor\xspace}
\newcommand{\Devin}{Devin\xspace}
\newcommand{\Jules}{Google Jules\xspace}
\newcommand{\Codex}{OpenAI Codex\xspace}

\newcommand{\AIDevVersion}{v5}

\newcommand{\AIDevCutoff}{March 31, 2026}
\newcommand{\MinStars}{100}
\newcommand{\NCandidateRepos}{12,994}
\newcommand{\NCreatedBeforeRepos}{5,550}
\newcommand{\NScreenedRepos}{1,227}
\newcommand{\NCommonRepos}{1,226}
\newcommand{\MinCommitsPerEra}{400}
\newcommand{\WindowMonths}{15}
\newcommand{\PreWindowStart}{2020-03-29}

\newcommand{\LLMWindowStart}{2023-09-24}

\newcommand{\AgentWindowStart}{2025-06-01}
\newcommand{\PreWindowLastDate}{2021-06-28}
\newcommand{\LLMWindowLastDate}{2024-12-23}
\newcommand{\AgentWindowLastDate}{2026-08-31}
\newcommand{\CreatedBefore}{\PreWindowStart} 

\newcommand{\CopilotAgentReleaseDate}{May 19, 2025} 
\newcommand{\CopilotPreviewLongDate}{June 29, 2021} 
\newcommand{\JulesBetaDate}{May 20, 2025} 
\newcommand{\NMultiAgentCommits}{243}

\newcommand{\NClaudeCode}{\TabDatasetAgentCCCommits}

\newcommand{\NStudyCommits}{\TabDatasetTotalCommits}
\newcommand{\NFeatures}{39}
\newcommand{\NReducedFeatures}{32} 
\newcommand{\NMsgFeatures}{10}
\newcommand{\NFileFeatures}{8}
\newcommand{\NDiffFeatures}{4}
\newcommand{\NStyleFeatures}{13}
\newcommand{\NLexFeatures}{4}
\newcommand{\CorrThreshold}{0.70}
\newcommand{\NFolds}{five}
\newcommand{\NBootstrap}{1,000}
\newcommand{\AgentShare}{0.059} 

\newcommand{\UnmaskedAUC}{0.859} 

\newcommand{\AUCXGB}{\TabModelsFullXGBAUC} 

\newcommand{\PrecXGB}{\TabModelsFullXGBPrec}

\newcommand{\AUCTFIDF}{\TabModelsFullTFIDFAUC} 

\newcommand{\PrecTFIDF}{\TabModelsFullTFIDFPrec}

\newcommand{\AUCDiffOnly}{\TabModelsFullTFIDFDiffAUC} 
\newcommand{\AUCNoMessage}{0.645} 
\newcommand{\AUCSizeOnly}{0.789} 
\newcommand{\AUCXGBSample}{\TabModelsSampleXGBAUC} 

\newcommand{\AUCSolFew}{\TabModelsSampleSolFewAUC}

\newcommand{\AUCMatchedCCXGB}{\TabAgentsCCMatchedAUC} 
\newcommand{\AUCMatchedCCTFIDF}{0.600}
\newcommand{\AUCMatchedCopXGB}{\TabAgentsCopMatchedAUC}
\newcommand{\AUCMatchedCopTFIDF}{0.737}
\newcommand{\AUCMatchedCopWorkflow}{0.907} 
\newcommand{\TopOnePctPrec}{0.316} 
\newcommand{\TopOnePctRec}{0.053}

\newcommand{\WFOneTFIDFFull}{\TabModelsFullTFIDFWFOne} 
\newcommand{\AUCCodeBERT}{\TabModelsSampleCodeBERTAUC} 
\newcommand{\AUCCodeBERTCode}{\TabModelsSampleCodeBERTDiffAUC} 
\newcommand{\PrecCodeBERT}{\TabModelsSampleCodeBERTPrec} 
\newcommand{\NCodeBERTTrain}{100,000} 
\newcommand{\WFOneAgenTag}{0.769} 

\newcommand{\AUCUnseenCC}{\TabUnseenCCXGBAUC} 
\newcommand{\AUCUnseenCop}{\TabUnseenCopXGBAUC}
\newcommand{\AUCUnseenCCTFIDF}{\TabUnseenCCTFIDFAUC} 
\newcommand{\AUCUnseenCopTFIDF}{\TabUnseenCopTFIDFAUC}
\newcommand{\RecUnseenCC}{\TabUnseenCCXGBRec} 
\newcommand{\AUCCopFirst}{0.83} 
\newcommand{\AUCCopLow}{0.61} 
\newcommand{\AUCCopLast}{0.78} 

\newcommand{\AUCTFIDFFrozenFirst}{0.733} 
\newcommand{\AUCTFIDFFrozenLast}{0.631}
\newcommand{\AUCTFIDFRetrained}{0.801}

\newcommand{\MsgWordsCC}{73} 
\newcommand{\MsgWordsHuman}{10}

\newcommand{\NSample}{4,000} 
\newcommand{\NSampleAgent}{2,000}
\newcommand{\NSampleHuman}{2,000}
\newcommand{\NFewShot}{16} 
\newcommand{\NPriorFeatures}{12} 
\newcommand{\NNewFeatures}{27} 
\newcommand{\VIFThreshold}{5}
\newcommand{\DefaultThreshold}{0.500} 
\newcommand{\RandomAUC}{0.500}
\newcommand{\MinEvalCommits}{20} 
\newcommand{\MinEvalRepos}{two} 
\newcommand{\FirstAgentPRDate}{December 24, 2024} 
\newcommand{\FirstMonth}{June 2025} 
\newcommand{\FirstTestMonth}{July 2025} 
\newcommand{\LastTrainMonth}{July 2026} 
\newcommand{\LastMonth}{August 2026} 

\newcommand{\AgenTagHumanMarkerPct}{5.0} 
\newcommand{\NAgenTagHumanPRs}{6,618} 
\newcommand{\ClaudePRRevisedPct}{45.1} 
\newcommand{\NFingerprintAgents}{five} 
\newcommand{\MotivFiles}{95} 
\newcommand{\MotivLines}{8,770} 
\newcommand{\MotivEndorsements}{240} 
\newcommand{\MinRQOneCommits}{20} 
\newcommand{\CliffNegligible}{0.147} 
\newcommand{\CliffSmall}{0.330}
\newcommand{\CliffMedium}{0.474}
\newcommand{\NBootstrapRQOne}{500}
\newcommand{\MinWilcoxonRepos}{five}

\newcommand{\DeltaDisplayCutoff}{0.100} 
\newcommand{\NReposRQOneCC}{461} 
\newcommand{\NReposRQOneCop}{202}
\newcommand{\NReposRQOneCursor}{13}

\newcommand{\NHumanEraRepos}{\TabDatasetPreLLMHumanRepos} 

\newcommand{\SizeBinsWords}{0, 1--10, 11--50, 51--200, and more than 200} 
\newcommand{\SizeBinsLines}{0, 1--10, 11--100, 101--1,000, and more than 1,000}
\newcommand{\SizeBinsFiles}{0, 1, 2--5, 6--20, and more than 20}
\newcommand{\BudgetSmall}{1} 
\newcommand{\BudgetMid}{5}
\newcommand{\BudgetLarge}{10}

\newcommand{\NRQTwoTest}{2,704,768} 
\newcommand{\NRQTwoHuman}{2,544,682}
\newcommand{\PrecXGBCC}{\TabAgentsCCPrec}

\newcommand{\AUCAgentMax}{\TabAgentsDevinAUC} 
\newcommand{\LiftAgentMin}{\TabAgentsCCLift} 
\newcommand{\LiftAgentMax}{\TabAgentsDevinLift}
\newcommand{\PrecSmallAgentsMax}{\TabAgentsCursorPrec} 
\newcommand{\ShareSmallAgentsMax}{0.001} 
\newcommand{\AUCAblationMin}{0.786} 
\newcommand{\AUCAblationMax}{0.795}

\newcommand{\AUCXGBCC}{\TabAgentsCCAUC} 
\newcommand{\AUCTFIDFCC}{0.860}

\newcommand{\AUCMatchedCCDiff}{0.533} 
\newcommand{\AUCMatchedCopDiff}{0.545}

\newcommand{\CoauthorCopShare}{0.852} 
\newcommand{\CoauthorHumanShare}{0.060} 
\newcommand{\AUCTFIDFAllLines}{0.884} 
\newcommand{\AUCNoAutoSpearman}{0.853}
\newcommand{\AUCLightGBM}{\TabModelsFullLGBMAUC}
\newcommand{\AUCCatBoost}{\TabModelsFullCatBoostAUC}
\newcommand{\AUCEqualWeight}{0.813} 

\newcommand{\PatchGain}{0.013} 

\newcommand{\AUCLLMEraHuman}{0.841} 

\newcommand{\NMinAgentTrain}{\TabTrainingCCTrainAgent} 
\newcommand{\NMinHumanTrain}{\TabTrainingCCTrainHuman} 
\newcommand{\AUCUnseenSmallMin}{\TabUnseenDevinXGBAUC} 
\newcommand{\AUCUnseenJules}{\TabUnseenJulesXGBAUC}

\newcommand{\RecUnseenCCDefault}{0.123} 
\newcommand{\PrecUnseenJules}{\TabUnseenJulesXGBPrec}
\newcommand{\AUCUnseenCCRelaxed}{0.772} 

\newcommand{\PrecUnseenMax}{0.070} 
\newcommand{\NMonthPairs}{105} 
\newcommand{\NTestMonths}{14} 
\newcommand{\CopLowMonth}{May 2026}
\newcommand{\AUCCCFrozenMin}{0.67}
\newcommand{\AUCCCFrozenMax}{0.78}

\newcommand{\InitialPlanHighMonth}{February 2026}
\newcommand{\InitialPlanLowMonth}{April 2026}
\newcommand{\InitialPlanShareHigh}{26.3\%} 
\newcommand{\InitialPlanShareLow}{8.8\%}

\newcommand{\RecCCFrozenMax}{0.091}
\newcommand{\PrecCCFrozenFirst}{0.028}
\newcommand{\PrecCCFrozenLast}{\TabFinalCCFirstPrec}
\newcommand{\ShareCCFirst}{0.003} 
\newcommand{\ShareCCLast}{0.108}
\newcommand{\ShareCopLast}{0.006}

\newcommand{\CopShiftMonth}{March 2026}

\newcommand{\AUCCCFinalFrozen}{\TabFinalCCFirstAUC} 
\newcommand{\AUCCCFinalRetrained}{\TabFinalCCLastAUC}

\newcommand{\NRQFourFinalHuman}{185,832} 
\newcommand{\NRQFourFinalCC}{22,545}
\newcommand{\NRQFourFinalCop}{1,101}
\newcommand{\NMonthsTFIDFBelow}{11} 

\newcommand{\CopDropMonths}{ten} 

\newcommand{\AUCCodeBERTMessage}{0.852} 

\newcommand{\WFOneCodeBERTMessage}{0.780} 

\newcommand{\TabDatasetPreLLMHumanRepos}{1,224} 
\newcommand{\TabDatasetPreLLMHumanCommits}{2,504,139} 
\newcommand{\TabDatasetPreLLMHumanMedianWords}{7} 
\newcommand{\TabDatasetPreLLMHumanFiles}{15,795,739} 
\newcommand{\TabDatasetPreLLMHumanLOC}{1,715,849,223} 
\newcommand{\TabDatasetPreLLMHumanMedianLOC}{16} 
\newcommand{\TabDatasetLLMHumanRepos}{1,224} 
\newcommand{\TabDatasetLLMHumanCommits}{2,509,950} 
\newcommand{\TabDatasetLLMHumanMedianWords}{8} 
\newcommand{\TabDatasetLLMHumanFiles}{18,206,607} 
\newcommand{\TabDatasetLLMHumanLOC}{1,559,030,746} 
\newcommand{\TabDatasetLLMHumanMedianLOC}{16} 
\newcommand{\TabDatasetAgentCopRepos}{697} 
\newcommand{\TabDatasetAgentCopCommits}{27,179} 
\newcommand{\TabDatasetAgentCopMedianWords}{14} 
\newcommand{\TabDatasetAgentCopFiles}{190,486} 
\newcommand{\TabDatasetAgentCopLOC}{21,634,462} 
\newcommand{\TabDatasetAgentCopMedianLOC}{26} 
\newcommand{\TabDatasetAgentCCRepos}{906} 
\newcommand{\TabDatasetAgentCCCommits}{141,034} 
\newcommand{\TabDatasetAgentCCMedianWords}{65} 
\newcommand{\TabDatasetAgentCCFiles}{1,147,521} 
\newcommand{\TabDatasetAgentCCLOC}{89,580,602} 
\newcommand{\TabDatasetAgentCCMedianLOC}{60} 
\newcommand{\TabDatasetAgentCursorRepos}{27} 
\newcommand{\TabDatasetAgentCursorCommits}{1,160} 
\newcommand{\TabDatasetAgentCursorMedianWords}{36} 
\newcommand{\TabDatasetAgentCursorFiles}{4,855} 
\newcommand{\TabDatasetAgentCursorLOC}{280,170} 
\newcommand{\TabDatasetAgentCursorMedianLOC}{41} 
\newcommand{\TabDatasetAgentDevinRepos}{10} 
\newcommand{\TabDatasetAgentDevinCommits}{75} 
\newcommand{\TabDatasetAgentDevinMedianWords}{60} 
\newcommand{\TabDatasetAgentDevinFiles}{156} 
\newcommand{\TabDatasetAgentDevinLOC}{6,755} 
\newcommand{\TabDatasetAgentDevinMedianLOC}{47} 
\newcommand{\TabDatasetAgentJulesRepos}{25} 
\newcommand{\TabDatasetAgentJulesCommits}{313} 
\newcommand{\TabDatasetAgentJulesMedianWords}{56} 
\newcommand{\TabDatasetAgentJulesFiles}{1,527} 
\newcommand{\TabDatasetAgentJulesLOC}{81,893} 
\newcommand{\TabDatasetAgentJulesMedianLOC}{36} 
\newcommand{\TabDatasetAgentHumanRepos}{1,226} 
\newcommand{\TabDatasetAgentHumanCommits}{2,654,096} 
\newcommand{\TabDatasetAgentHumanMedianWords}{9} 
\newcommand{\TabDatasetAgentHumanFiles}{19,289,038} 
\newcommand{\TabDatasetAgentHumanLOC}{1,911,987,960} 
\newcommand{\TabDatasetAgentHumanMedianLOC}{21} 
\newcommand{\TabDatasetTotalCommits}{7,837,946} 
\newcommand{\TabDatasetTotalFiles}{54,635,929} 
\newcommand{\TabDatasetTotalLOC}{5,298,451,811} 

\newcommand{\TabEffectsCharactersCC}{0.698} 
\newcommand{\TabEffectsCharactersCop}{0.508} 
\newcommand{\TabEffectsCharactersCursor}{0.764} 
\newcommand{\TabEffectsWordsCC}{0.693} 
\newcommand{\TabEffectsWordsCop}{0.422} 
\newcommand{\TabEffectsWordsCursor}{0.692} 
\newcommand{\TabEffectsLinesCC}{0.641} 
\newcommand{\TabEffectsLinesCop}{0.544} 
\newcommand{\TabEffectsLinesCursor}{0.786} 
\newcommand{\TabEffectsConventionalPrefixCC}{0.021} 
\newcommand{\TabEffectsConventionalPrefixCop}{0.000} 
\newcommand{\TabEffectsConventionalPrefixCursor}{0.144} 
\newcommand{\TabEffectsAdditionShareCC}{0.150} 
\newcommand{\TabEffectsAdditionShareCop}{0.123} 
\newcommand{\TabEffectsAdditionShareCursor}{0.153} 
\newcommand{\TabEffectsChangedLinesPerFileCC}{0.193} 
\newcommand{\TabEffectsChangedLinesPerFileCop}{0.063} 
\newcommand{\TabEffectsChangedLinesPerFileCursor}{0.089} 
\newcommand{\TabEffectsSingleFileCommitCC}{-0.084} 
\newcommand{\TabEffectsSingleFileCommitCop}{-0.102} 
\newcommand{\TabEffectsSingleFileCommitCursor}{0.016} 
\newcommand{\TabEffectsAddedLinesInPatchesCC}{0.218} 
\newcommand{\TabEffectsAddedLinesInPatchesCop}{0.108} 
\newcommand{\TabEffectsAddedLinesInPatchesCursor}{0.094} 
\newcommand{\TabEffectsIdentifiersCC}{0.241} 
\newcommand{\TabEffectsIdentifiersCop}{0.119} 
\newcommand{\TabEffectsIdentifiersCursor}{0.100} 
\newcommand{\TabEffectsCamelCaseShareCC}{0.036} 
\newcommand{\TabEffectsCamelCaseShareCop}{0.011} 
\newcommand{\TabEffectsCamelCaseShareCursor}{0.109} 
\newcommand{\TabEffectsIndentationVarianceCC}{0.115} 
\newcommand{\TabEffectsIndentationVarianceCop}{0.045} 
\newcommand{\TabEffectsIndentationVarianceCursor}{0.032} 
\newcommand{\TabEffectsTokensCC}{0.237} 
\newcommand{\TabEffectsTokensCop}{0.110} 
\newcommand{\TabEffectsTokensCursor}{0.092} 
\newcommand{\TabEffectsDistinctTokensCC}{0.264} 
\newcommand{\TabEffectsDistinctTokensCop}{0.101} 
\newcommand{\TabEffectsDistinctTokensCursor}{0.087} 

\newcommand{\TabErasCharactersPreToLLM}{0.399} 
\newcommand{\TabErasCharactersLLMToAgent}{0.563} 
\newcommand{\TabErasCharactersPreToAgent}{0.663} 
\newcommand{\TabErasWordsPreToLLM}{0.372} 
\newcommand{\TabErasWordsLLMToAgent}{0.567} 
\newcommand{\TabErasWordsPreToAgent}{0.629} 
\newcommand{\TabErasLinesPreToLLM}{0.509} 
\newcommand{\TabErasLinesLLMToAgent}{0.575} 
\newcommand{\TabErasLinesPreToAgent}{0.689} 
\newcommand{\TabErasConventionalPrefixPreToLLM}{0.572} 
\newcommand{\TabErasConventionalPrefixLLMToAgent}{0.680} 
\newcommand{\TabErasConventionalPrefixPreToAgent}{0.827} 
\newcommand{\TabErasAdditionSharePreToLLM}{0.009} 
\newcommand{\TabErasAdditionShareLLMToAgent}{0.632} 
\newcommand{\TabErasAdditionSharePreToAgent}{0.524} 
\newcommand{\TabErasChangedLinesPerFilePreToLLM}{0.030} 
\newcommand{\TabErasChangedLinesPerFileLLMToAgent}{0.738} 
\newcommand{\TabErasChangedLinesPerFilePreToAgent}{0.628} 
\newcommand{\TabErasSingleFileCommitPreToLLM}{-0.233} 
\newcommand{\TabErasSingleFileCommitLLMToAgent}{-0.417} 
\newcommand{\TabErasSingleFileCommitPreToAgent}{-0.483} 
\newcommand{\TabErasAddedLinesInPatchesPreToLLM}{0.127} 
\newcommand{\TabErasAddedLinesInPatchesLLMToAgent}{0.693} 
\newcommand{\TabErasAddedLinesInPatchesPreToAgent}{0.629} 
\newcommand{\TabErasIdentifiersPreToLLM}{0.127} 
\newcommand{\TabErasIdentifiersLLMToAgent}{0.704} 
\newcommand{\TabErasIdentifiersPreToAgent}{0.651} 
\newcommand{\TabErasCamelCaseSharePreToLLM}{0.086} 
\newcommand{\TabErasCamelCaseShareLLMToAgent}{0.097} 
\newcommand{\TabErasCamelCaseSharePreToAgent}{0.119} 
\newcommand{\TabErasIndentationVariancePreToLLM}{0.029} 
\newcommand{\TabErasIndentationVarianceLLMToAgent}{0.407} 
\newcommand{\TabErasIndentationVariancePreToAgent}{0.318} 
\newcommand{\TabErasTokensPreToLLM}{0.134} 
\newcommand{\TabErasTokensLLMToAgent}{0.697} 
\newcommand{\TabErasTokensPreToAgent}{0.658} 
\newcommand{\TabErasDistinctTokensPreToLLM}{0.104} 
\newcommand{\TabErasDistinctTokensLLMToAgent}{0.737} 
\newcommand{\TabErasDistinctTokensPreToAgent}{0.670} 

\newcommand{\TabModelsFullXGBAUC}{0.792} 
\newcommand{\TabModelsFullXGBPRAUC}{0.187} 
\newcommand{\TabModelsFullXGBLift}{3.152} 
\newcommand{\TabModelsFullXGBPrec}{0.184} 
\newcommand{\TabModelsFullXGBRec}{0.473} 
\newcommand{\TabModelsFullXGBMCC}{0.224} 
\newcommand{\TabModelsFullXGBWFOne}{0.875} 
\newcommand{\TabModelsFullLGBMAUC}{0.792} 
\newcommand{\TabModelsFullLGBMPRAUC}{0.187} 
\newcommand{\TabModelsFullLGBMLift}{3.161} 
\newcommand{\TabModelsFullLGBMPrec}{0.182} 
\newcommand{\TabModelsFullLGBMRec}{0.476} 
\newcommand{\TabModelsFullLGBMMCC}{0.223} 
\newcommand{\TabModelsFullLGBMWFOne}{0.874} 
\newcommand{\TabModelsFullCatBoostAUC}{0.790} 
\newcommand{\TabModelsFullCatBoostPRAUC}{0.188} 
\newcommand{\TabModelsFullCatBoostLift}{3.175} 
\newcommand{\TabModelsFullCatBoostPrec}{0.186} 
\newcommand{\TabModelsFullCatBoostRec}{0.461} 
\newcommand{\TabModelsFullCatBoostMCC}{0.223} 
\newcommand{\TabModelsFullCatBoostWFOne}{0.877} 
\newcommand{\TabModelsFullTFIDFAUC}{0.859} 
\newcommand{\TabModelsFullTFIDFPRAUC}{0.305} 
\newcommand{\TabModelsFullTFIDFLift}{5.151} 
\newcommand{\TabModelsFullTFIDFPrec}{0.332} 
\newcommand{\TabModelsFullTFIDFRec}{0.497} 
\newcommand{\TabModelsFullTFIDFMCC}{0.361} 
\newcommand{\TabModelsFullTFIDFWFOne}{0.919} 
\newcommand{\TabModelsFullTFIDFMsgAUC}{0.859} 
\newcommand{\TabModelsFullTFIDFMsgPRAUC}{0.314} 
\newcommand{\TabModelsFullTFIDFMsgLift}{5.306} 
\newcommand{\TabModelsFullTFIDFMsgPrec}{0.320} 
\newcommand{\TabModelsFullTFIDFMsgRec}{0.534} 
\newcommand{\TabModelsFullTFIDFMsgMCC}{0.366} 
\newcommand{\TabModelsFullTFIDFMsgWFOne}{0.916} 
\newcommand{\TabModelsFullTFIDFDiffAUC}{0.716} 
\newcommand{\TabModelsFullTFIDFDiffPRAUC}{0.136} 
\newcommand{\TabModelsFullTFIDFDiffLift}{2.299} 
\newcommand{\TabModelsFullTFIDFDiffPrec}{0.146} 
\newcommand{\TabModelsFullTFIDFDiffRec}{0.388} 
\newcommand{\TabModelsFullTFIDFDiffMCC}{0.158} 
\newcommand{\TabModelsFullTFIDFDiffWFOne}{0.863} 
\newcommand{\TabModelsSampleXGBAUC}{0.783} 
\newcommand{\TabModelsSampleXGBPRAUC}{0.740} 
\newcommand{\TabModelsSampleXGBLift}{1.480} 
\newcommand{\TabModelsSampleXGBPrec}{0.160} 
\newcommand{\TabModelsSampleXGBRec}{0.510} 
\newcommand{\TabModelsSampleXGBMCC}{0.206} 
\newcommand{\TabModelsSampleXGBWFOne}{0.662} 
\newcommand{\TabModelsSampleLGBMAUC}{0.786} 
\newcommand{\TabModelsSampleLGBMPRAUC}{0.751} 
\newcommand{\TabModelsSampleLGBMLift}{1.502} 
\newcommand{\TabModelsSampleLGBMPrec}{0.157} 
\newcommand{\TabModelsSampleLGBMRec}{0.512} 
\newcommand{\TabModelsSampleLGBMMCC}{0.203} 
\newcommand{\TabModelsSampleLGBMWFOne}{0.661} 
\newcommand{\TabModelsSampleCatBoostAUC}{0.783} 
\newcommand{\TabModelsSampleCatBoostPRAUC}{0.745} 
\newcommand{\TabModelsSampleCatBoostLift}{1.489} 
\newcommand{\TabModelsSampleCatBoostPrec}{0.162} 
\newcommand{\TabModelsSampleCatBoostRec}{0.506} 
\newcommand{\TabModelsSampleCatBoostMCC}{0.207} 
\newcommand{\TabModelsSampleCatBoostWFOne}{0.662} 
\newcommand{\TabModelsSampleTFIDFAUC}{0.799} 
\newcommand{\TabModelsSampleTFIDFPRAUC}{0.786} 
\newcommand{\TabModelsSampleTFIDFLift}{1.573} 
\newcommand{\TabModelsSampleTFIDFPrec}{0.246} 
\newcommand{\TabModelsSampleTFIDFRec}{0.463} 
\newcommand{\TabModelsSampleTFIDFMCC}{0.280} 
\newcommand{\TabModelsSampleTFIDFWFOne}{0.670} 
\newcommand{\TabModelsSampleTFIDFMsgAUC}{0.807} 
\newcommand{\TabModelsSampleTFIDFMsgPRAUC}{0.794} 
\newcommand{\TabModelsSampleTFIDFMsgLift}{1.588} 
\newcommand{\TabModelsSampleTFIDFMsgPrec}{0.220} 
\newcommand{\TabModelsSampleTFIDFMsgRec}{0.501} 
\newcommand{\TabModelsSampleTFIDFMsgMCC}{0.269} 
\newcommand{\TabModelsSampleTFIDFMsgWFOne}{0.683} 
\newcommand{\TabModelsSampleTFIDFDiffAUC}{0.621} 
\newcommand{\TabModelsSampleTFIDFDiffPRAUC}{0.607} 
\newcommand{\TabModelsSampleTFIDFDiffLift}{1.215} 
\newcommand{\TabModelsSampleTFIDFDiffPrec}{0.095} 
\newcommand{\TabModelsSampleTFIDFDiffRec}{0.359} 
\newcommand{\TabModelsSampleTFIDFDiffMCC}{0.082} 
\newcommand{\TabModelsSampleTFIDFDiffWFOne}{0.552} 
\newcommand{\TabModelsSampleCodeBERTAUC}{0.846} 
\newcommand{\TabModelsSampleCodeBERTPRAUC}{0.822} 
\newcommand{\TabModelsSampleCodeBERTLift}{1.644} 
\newcommand{\TabModelsSampleCodeBERTPrec}{0.169} 
\newcommand{\TabModelsSampleCodeBERTRec}{0.797} 
\newcommand{\TabModelsSampleCodeBERTMCC}{0.289} 
\newcommand{\TabModelsSampleCodeBERTWFOne}{0.775} 
\newcommand{\TabModelsSampleCodeBERTMsgAUC}{0.852} 
\newcommand{\TabModelsSampleCodeBERTMsgPRAUC}{0.829} 
\newcommand{\TabModelsSampleCodeBERTMsgLift}{1.658} 
\newcommand{\TabModelsSampleCodeBERTMsgPrec}{0.169} 
\newcommand{\TabModelsSampleCodeBERTMsgRec}{0.809} 
\newcommand{\TabModelsSampleCodeBERTMsgMCC}{0.293} 
\newcommand{\TabModelsSampleCodeBERTMsgWFOne}{0.780} 
\newcommand{\TabModelsSampleCodeBERTDiffAUC}{0.589} 
\newcommand{\TabModelsSampleCodeBERTDiffPRAUC}{0.571} 
\newcommand{\TabModelsSampleCodeBERTDiffLift}{1.142} 
\newcommand{\TabModelsSampleCodeBERTDiffPrec}{0.063} 
\newcommand{\TabModelsSampleCodeBERTDiffRec}{0.906} 
\newcommand{\TabModelsSampleCodeBERTDiffMCC}{0.042} 
\newcommand{\TabModelsSampleCodeBERTDiffWFOne}{0.455} 
\newcommand{\TabModelsSampleLunaZeroAUC}{0.664} 
\newcommand{\TabModelsSampleLunaZeroPRAUC}{0.650} 
\newcommand{\TabModelsSampleLunaZeroLift}{1.300} 
\newcommand{\TabModelsSampleLunaZeroPrec}{0.106} 
\newcommand{\TabModelsSampleLunaZeroRec}{0.564} 
\newcommand{\TabModelsSampleLunaZeroMCC}{0.135} 
\newcommand{\TabModelsSampleLunaZeroWFOne}{0.631} 
\newcommand{\TabModelsSampleSolZeroAUC}{0.797} 
\newcommand{\TabModelsSampleSolZeroPRAUC}{0.757} 
\newcommand{\TabModelsSampleSolZeroLift}{1.513} 
\newcommand{\TabModelsSampleSolZeroPrec}{0.154} 
\newcommand{\TabModelsSampleSolZeroRec}{0.710} 
\newcommand{\TabModelsSampleSolZeroMCC}{0.247} 
\newcommand{\TabModelsSampleSolZeroWFOne}{0.732} 
\newcommand{\TabModelsSampleSolFewAUC}{0.802} 
\newcommand{\TabModelsSampleSolFewPRAUC}{0.761} 
\newcommand{\TabModelsSampleSolFewLift}{1.522} 
\newcommand{\TabModelsSampleSolFewPrec}{0.143} 
\newcommand{\TabModelsSampleSolFewRec}{0.792} 
\newcommand{\TabModelsSampleSolFewMCC}{0.247} 
\newcommand{\TabModelsSampleSolFewWFOne}{0.746} 

\newcommand{\TabAgentsCCShare}{0.050} 
\newcommand{\TabAgentsCCAUC}{0.786} 
\newcommand{\TabAgentsCCPRAUC}{0.157} 
\newcommand{\TabAgentsCCLift}{3.123} 
\newcommand{\TabAgentsCCPrec}{0.156} 
\newcommand{\TabAgentsCCRec}{0.458} 
\newcommand{\TabAgentsCCMCC}{0.201} 
\newcommand{\TabAgentsCCMatchedAUC}{0.543} 
\newcommand{\TabAgentsCopShare}{0.009} 
\newcommand{\TabAgentsCopAUC}{0.824} 
\newcommand{\TabAgentsCopPRAUC}{0.040} 
\newcommand{\TabAgentsCopLift}{4.390} 
\newcommand{\TabAgentsCopPrec}{0.037} 
\newcommand{\TabAgentsCopRec}{0.549} 
\newcommand{\TabAgentsCopMCC}{0.116} 
\newcommand{\TabAgentsCopMatchedAUC}{0.692} 
\newcommand{\TabAgentsCursorShare}{$<$0.001} 
\newcommand{\TabAgentsCursorAUC}{0.879} 
\newcommand{\TabAgentsCursorPRAUC}{0.004} 
\newcommand{\TabAgentsCursorLift}{8.292} 
\newcommand{\TabAgentsCursorPrec}{0.002} 
\newcommand{\TabAgentsCursorRec}{0.648} 
\newcommand{\TabAgentsCursorMCC}{0.032} 
\newcommand{\TabAgentsCursorMatchedAUC}{0.655} 
\newcommand{\TabAgentsDevinShare}{$<$0.001} 
\newcommand{\TabAgentsDevinAUC}{0.904} 
\newcommand{\TabAgentsDevinPRAUC}{$<$0.001} 
\newcommand{\TabAgentsDevinLift}{15.121} 
\newcommand{\TabAgentsDevinPrec}{$<$0.001} 
\newcommand{\TabAgentsDevinRec}{0.730} 
\newcommand{\TabAgentsDevinMCC}{0.010} 
\newcommand{\TabAgentsDevinMatchedAUC}{0.703} 
\newcommand{\TabAgentsJulesShare}{$<$0.001} 
\newcommand{\TabAgentsJulesAUC}{0.853} 
\newcommand{\TabAgentsJulesPRAUC}{0.001} 
\newcommand{\TabAgentsJulesLift}{5.429} 
\newcommand{\TabAgentsJulesPrec}{0.001} 
\newcommand{\TabAgentsJulesRec}{0.623} 
\newcommand{\TabAgentsJulesMCC}{0.016} 
\newcommand{\TabAgentsJulesMatchedAUC}{0.675} 

\newcommand{\TabTrainingCCTrainAgent}{6,527} 
\newcommand{\TabTrainingCCTrainHuman}{446,694} 
\newcommand{\TabTrainingCCTestAgent}{141,034} 
\newcommand{\TabTrainingCCTestHuman}{2,207,402} 
\newcommand{\TabTrainingCopTrainAgent}{53,251} 
\newcommand{\TabTrainingCopTrainHuman}{1,139,036} 
\newcommand{\TabTrainingCopTestAgent}{27,179} 
\newcommand{\TabTrainingCopTestHuman}{1,515,060} 
\newcommand{\TabTrainingCursorTrainAgent}{146,865} 
\newcommand{\TabTrainingCursorTrainHuman}{2,543,735} 
\newcommand{\TabTrainingCursorTestAgent}{1,160} 
\newcommand{\TabTrainingCursorTestHuman}{110,361} 
\newcommand{\TabTrainingDevinTrainAgent}{168,366} 
\newcommand{\TabTrainingDevinTrainHuman}{2,630,883} 
\newcommand{\TabTrainingDevinTestAgent}{75} 
\newcommand{\TabTrainingDevinTestHuman}{23,213} 
\newcommand{\TabTrainingJulesTrainAgent}{166,019} 
\newcommand{\TabTrainingJulesTrainHuman}{2,526,866} 
\newcommand{\TabTrainingJulesTestAgent}{313} 
\newcommand{\TabTrainingJulesTestHuman}{127,230} 

\newcommand{\TabUnseenCCShare}{0.060} 
\newcommand{\TabUnseenCCXGBAUC}{0.702} 
\newcommand{\TabUnseenCCXGBPRAUC}{0.109} 
\newcommand{\TabUnseenCCXGBLift}{1.809} 
\newcommand{\TabUnseenCCXGBPrec}{0.069} 
\newcommand{\TabUnseenCCXGBRec}{0.005} 
\newcommand{\TabUnseenCCXGBMCC}{0.003} 
\newcommand{\TabUnseenCCTFIDFAUC}{0.524} 
\newcommand{\TabUnseenCCTFIDFPRAUC}{0.068} 
\newcommand{\TabUnseenCCTFIDFLift}{1.131} 
\newcommand{\TabUnseenCCTFIDFPrec}{0.086} 
\newcommand{\TabUnseenCCTFIDFRec}{0.024} 
\newcommand{\TabUnseenCCTFIDFMCC}{0.014} 
\newcommand{\TabUnseenCopShare}{0.018} 
\newcommand{\TabUnseenCopXGBAUC}{0.706} 
\newcommand{\TabUnseenCopXGBPRAUC}{0.039} 
\newcommand{\TabUnseenCopXGBLift}{2.209} 
\newcommand{\TabUnseenCopXGBPrec}{0.047} 
\newcommand{\TabUnseenCopXGBRec}{0.250} 
\newcommand{\TabUnseenCopXGBMCC}{0.072} 
\newcommand{\TabUnseenCopTFIDFAUC}{0.630} 
\newcommand{\TabUnseenCopTFIDFPRAUC}{0.028} 
\newcommand{\TabUnseenCopTFIDFLift}{1.608} 
\newcommand{\TabUnseenCopTFIDFPrec}{0.034} 
\newcommand{\TabUnseenCopTFIDFRec}{0.143} 
\newcommand{\TabUnseenCopTFIDFMCC}{0.036} 
\newcommand{\TabUnseenCursorShare}{0.010} 
\newcommand{\TabUnseenCursorXGBAUC}{0.898} 
\newcommand{\TabUnseenCursorXGBPRAUC}{0.067} 
\newcommand{\TabUnseenCursorXGBLift}{6.396} 
\newcommand{\TabUnseenCursorXGBPrec}{0.069} 
\newcommand{\TabUnseenCursorXGBRec}{0.552} 
\newcommand{\TabUnseenCursorXGBMCC}{0.174} 
\newcommand{\TabUnseenCursorTFIDFAUC}{0.739} 
\newcommand{\TabUnseenCursorTFIDFPRAUC}{0.023} 
\newcommand{\TabUnseenCursorTFIDFLift}{2.224} 
\newcommand{\TabUnseenCursorTFIDFPrec}{0.029} 
\newcommand{\TabUnseenCursorTFIDFRec}{0.293} 
\newcommand{\TabUnseenCursorTFIDFMCC}{0.064} 
\newcommand{\TabUnseenDevinShare}{0.003} 
\newcommand{\TabUnseenDevinXGBAUC}{0.893} 
\newcommand{\TabUnseenDevinXGBPRAUC}{0.028} 
\newcommand{\TabUnseenDevinXGBLift}{8.549} 
\newcommand{\TabUnseenDevinXGBPrec}{0.016} 
\newcommand{\TabUnseenDevinXGBRec}{0.707} 
\newcommand{\TabUnseenDevinXGBMCC}{0.094} 
\newcommand{\TabUnseenDevinTFIDFAUC}{0.889} 
\newcommand{\TabUnseenDevinTFIDFPRAUC}{0.030} 
\newcommand{\TabUnseenDevinTFIDFLift}{9.173} 
\newcommand{\TabUnseenDevinTFIDFPrec}{0.032} 
\newcommand{\TabUnseenDevinTFIDFRec}{0.560} 
\newcommand{\TabUnseenDevinTFIDFMCC}{0.123} 
\newcommand{\TabUnseenJulesShare}{0.002} 
\newcommand{\TabUnseenJulesXGBAUC}{0.921} 
\newcommand{\TabUnseenJulesXGBPRAUC}{0.022} 
\newcommand{\TabUnseenJulesXGBLift}{9.021} 
\newcommand{\TabUnseenJulesXGBPrec}{0.021} 
\newcommand{\TabUnseenJulesXGBRec}{0.546} 
\newcommand{\TabUnseenJulesXGBMCC}{0.097} 
\newcommand{\TabUnseenJulesTFIDFAUC}{0.835} 
\newcommand{\TabUnseenJulesTFIDFPRAUC}{0.012} 
\newcommand{\TabUnseenJulesTFIDFLift}{4.861} 
\newcommand{\TabUnseenJulesTFIDFPrec}{0.015} 
\newcommand{\TabUnseenJulesTFIDFRec}{0.137} 
\newcommand{\TabUnseenJulesTFIDFMCC}{0.039} 

\newcommand{\TabFinalCCFirstAUC}{0.705} 
\newcommand{\TabFinalCCFirstPRAUC}{0.222} 
\newcommand{\TabFinalCCFirstPrec}{0.320} 
\newcommand{\TabFinalCCFirstRec}{0.075} 
\newcommand{\TabFinalCCFirstFOne}{0.121} 
\newcommand{\TabFinalCCLastAUC}{0.752} 
\newcommand{\TabFinalCCLastPRAUC}{0.249} 
\newcommand{\TabFinalCCLastPrec}{0.312} 
\newcommand{\TabFinalCCLastRec}{0.185} 
\newcommand{\TabFinalCCLastFOne}{0.233} 
\newcommand{\TabFinalCopFirstAUC}{0.779} 
\newcommand{\TabFinalCopFirstPRAUC}{0.019} 
\newcommand{\TabFinalCopFirstPrec}{0.027} 
\newcommand{\TabFinalCopFirstRec}{0.092} 
\newcommand{\TabFinalCopFirstFOne}{0.042} 
\newcommand{\TabFinalCopLastAUC}{0.768} 
\newcommand{\TabFinalCopLastPRAUC}{0.015} 
\newcommand{\TabFinalCopLastPrec}{0.016} 
\newcommand{\TabFinalCopLastRec}{0.138} 
\newcommand{\TabFinalCopLastFOne}{0.029} 

\definecolor{darkgreen}{rgb}{0, 0.5, 0} 
\definecolor{whitesmoke}{rgb}{0.99, 0.99, 0.99} 

\def\fig#1{Fig.~\ref{#1}}
\def\tab#1{Table~\ref{#1}}

\def\sec#1{Section~\ref{#1}}
\def\Underline{\setbox0\hbox\bgroup\let\\\endUnderline}
\def\endUnderline{\vphantom{y}\egroup\smash{\underline{\box0}}\\}
\def\|{\verb|}

\newcommand{\etal}{\xspace\textit{et al.}\xspace}

\newcounter{findingctr}

\newcommand{\finding}[1]{%
  \smallskip
  \refstepcounter{findingctr}%
  \textbf{Finding~\#\thefindingctr: #1}%
}

\usepackage{listings}
\definecolor{backcolour}{rgb}{0.95,0.95,0.92}
\lstdefinelanguage{diff}{
  morecomment=**[f][\color{red}]{-},         
  morecomment=**[f][\color{darkgreen}]{+},       
  moredelim=**[is][\bfseries]{@@}{@@},
}
\definecolor{backcolour}{rgb}{0.95,0.95,0.92}
\lstdefinelanguage{commit}{ 
  breakindent = 0pt,
  numbers=none,
  backgroundcolor=\color{white},
  frame=single,
  xleftmargin=3.5em,
  numbersep=0em,
  xrightmargin=1.5em,
}

\usepackage[many]{tcolorbox}  
\tcbuselibrary{listings,breakable}
\definecolor{main}{HTML}{D0D3D4}    
\definecolor{sub}{HTML}{D0D3D4}     
\newtcolorbox{dbox}{
    left=2pt,right=2pt,top=2pt,bottom=2pt,
    enhanced, 
    boxrule = 0pt,
    enlarge top by=5pt,
    enlarge bottom by=3pt,
  }
\tcbset{
    sharp corners,
    before skip = 0.1cm,    
    after skip = 0.01cm      
}

\def\summarybox#1#2{
\medskip
\begin{tcolorbox}[
  enhanced,
  title=#1,
  colframe=darkgray,
]
    #2
\end{tcolorbox}
}

\usepackage{tcolorbox}
\tcbuselibrary{skins,breakable}
\newtcolorbox{reviewcomment}[1][Comment]{  
    enhanced,
    frame hidden,
    borderline west={4pt}{0pt}{gray!50},
    colback=gray!10,
    boxrule=0pt,
    sharp corners,
    breakable,
    before upper={\parindent0pt\textbf{#1:}\par\vspace{0.3em}},
    fontupper=\itshape,
    left=10pt,
    right=10pt,
    top=5pt,
    bottom=5pt,
    before skip=1.5\baselineskip
}

\AtBeginDocument{%
  \providecommand\BibTeX{{%
    \normalfont B\kern-0.5em{\scshape i\kern-0.25em b}\kern-0.8em\TeX}}}

\setcopyright{acmlicensed}
\copyrightyear{2027}
\acmYear{2027}
\acmDOI{XXXXXXX.XXXXXXX}

\begin{document}

\title{Agent or Human? Detecting Commits by AI Coding Agents in the Absence of Authorship Signatures}


\author{Kosei Horikawa}
\email{horikawa.kosei.hk1@naist.ac.jp}
\orcid{0009-0006-6317-0754}
\affiliation{%
  \institution{Nara Institute of Science and Technology}
  \city{Ikoma}
  \country{Japan}
}

\author{Hao Li}
\affiliation{%
  \institution{Queen's University}
  \city{Kingston}
  \country{Canada}
}
\email{hao.li@queensu.ca}
\orcid{0000-0003-4468-5972}

\author{Yutaro Kashiwa}
\affiliation{%
  \institution{Nara Institute of Science and Technology}
  \city{Ikoma}
  \country{Japan}
}
\email{yutaro.kashiwa@is.naist.jp}
\orcid{0000-0002-9633-7577}

\author{Youssef Esseddiq Ouatiti}
\affiliation{%
  \institution{Queen's University}
  \city{Kingston}
  \country{Canada}
}
\email{youssefesseddiq.ouatiti@queensu.ca}
\orcid{0000-0001-8063-6225}

\author{Mohammed Sayagh}
\affiliation{%
  \institution{École de technologie supérieure - Québec University}
  \city{Montréal}
  \country{Canada}
}
\email{mohammed.sayagh@etsmtl.ca}
\orcid{0000-0002-2724-0034}

\author{Bram Adams}
\affiliation{%
  \institution{Queen's University}
  \city{Kingston}
  \country{Canada}
}
\email{bram.adams@queensu.ca}
\orcid{0000-0001-7213-4006}

\author{Hajimu Iida}
\affiliation{%
  \institution{Nara Institute of Science and Technology}
  \city{Ikoma}
  \country{Japan}
}
\email{iida@itc.naist.jp}
\orcid{0000-0002-2919-6620}

\author{Ahmed E. Hassan}
\affiliation{%
  \institution{Queen's University}
  \city{Kingston}
  \country{Canada}
}
\email{ahmed@cs.queensu.ca}
\orcid{0000-0001-7749-5513}

\renewcommand{\shortauthors}{Horikawa, et al.}

\begin{abstract}
AI coding agents such as \ClaudeCode, \CopilotFull, and \Cursor now commit directly to open-source repositories.
Empirical studies of these agents identify their commits through authorship signatures, such as bot accounts and \texttt{Co-authored-by} trailers.
These signatures are optional, can be removed, and are often attached to pull requests rather than to individual commits.
Hence, it is unclear whether Agent commits can be told apart from Human commits once the signatures are gone.
We investigate the signals and limits of signature-free detection using \NStudyCommits{} commits from \NCommonRepos{} repositories, covering five coding agents and Human commits.
After signatures are masked, detectors rank Agent commits above Human commits with a ROC AUC of up to \AUCTFIDF{}, and most of this signal comes from message verbosity and change size rather than from the code itself.
On the shared sample, \texttt{gpt-6-sol} without fine-tuning reaches AUCs similar to XGBoost and TF-IDF.
At the observed Agent share, however, the best detector on the full test set reaches a precision of only \PrecTFIDF{}.
Detection transfers poorly to held-out \ClaudeCode and \Copilot, with TF-IDF reaching an AUC of only \AUCUnseenCCTFIDF{} for \ClaudeCode.
Temporal changes also depend on the agent, e.g., \Copilot AUC drops and later recovers, while more training history helps \ClaudeCode but not \Copilot.
These results indicate that signature-free detection can prioritize commits for review but cannot replace recorded attribution, and that studies relying on signature-based labels should treat them as a lower bound on agent activity.
\end{abstract}


\maketitle

\section{Introduction}\label{sec:introduction}

AI coding agents combine large language models with development tools to inspect repositories, edit files, run tests, and revise changes based on feedback~\cite{yang2024sweagent,anthropic2026claudecode,github2026copilotagent}.
Unlike code completion, an agent can complete a task from an issue to a pull request with little human input.
Agents such as \ClaudeCode, \CopilotShort, \Cursor, \Devin, and \Jules contribute to thousands of open-source projects, and their pull requests are studied for acceptance, testing, and maintenance~\cite{pinna2026acceptance,haque2026tests,sawada2026maintenance}.

Studying agent contributions requires knowing which changes an agent made.
Current work identifies agent contributions through authorship signatures: bot accounts, branch names, and attribution trailers such as \texttt{Co-Authored-By: Claude}~\cite{li2026aidev,robbes2026agentic}.
Whether a signature appears depends on tool settings.
\ClaudeCode lets users replace its commit attribution with any text~\cite{anthropic2026claudecodesettings}, \Cursor lets users and administrators turn its commit trailer off~\cite{cursor2026git}, and the default for adding \Copilot as a co-author in Visual Studio Code changed in three consecutive releases~\cite{microsoft2026vscodecoauthor}.
Signatures also attach to pull requests rather than individual commits, and human developers revise agent-authored code after merging~\cite{DBLP:journals/corr/abs-2509-14745,khelifi2026interventions}.
In \AIDev, \AgenTagHumanMarkerPct{}\% of human-labeled pull requests contain agent identifiers~\cite{ghaleb2026agentag}.

Recognizing Agent commits from their content could help researchers recover unsigned agent activity and check signature-based labels.
Without reliable content-based attribution, studies that rely on signatures measure only the agent use that developers choose to disclose.
Prior detection work classifies pull requests, either to identify the authoring agent~\cite{ghaleb2026fingerprinting} or to separate agent-authored from human-authored pull requests after removing agent identifiers~\cite{ghaleb2026agentag}.
A pull-request label, however, does not identify who wrote each commit, and commits reach the default branch without pull requests.
It remains unknown how well individual commits can be attributed once signatures are removed, at the prevalence at which agent commits actually occur, for agents that a detector has not seen, and as agent behavior changes over time.

We investigate whether Agent commits can be distinguished from Human commits after explicit authorship signatures are removed.
Starting from the \AIDev dataset~\cite{li2026aidev}, we collect \NStudyCommits{} default-branch commits from \NCommonRepos{} GitHub repositories.
We label commits from five agents by their signatures, and we compare them with Human commits from the same repositories in three eras: before LLM assistants, after LLM assistants, and after coding agents became available.
We then mask the signatures and evaluate detectors that separate Agent from Human commits in repositories absent from training.
Our study answers four research questions~(RQs):

\begin{itemize}[leftmargin=*]
  \item \textbf{\rqa: \textit{\rqaa}}
  To identify which characteristics a detector could use, we compare Agent and Human commits within the same repositories and era.
  We find that Agent commits differ mainly in message length.
  The repository-level median \ClaudeCode message has \MsgWordsCC{} words, compared with \MsgWordsHuman{} for Human commits, and Human commits themselves grew longer across eras.

  \item \textbf{\rqb: \textit{\rqbb}}
  We train detectors on masked commits and test them on unseen repositories at the observed Agent share of \AgentShare{}.
  Detectors rank Agent commits above Human commits with an ROC AUC of up to \AUCTFIDF{}, but at thresholds selected on validation data their precision is at most \PrecTFIDF{}.
  The signal comes mostly from the commit message and change size rather than from the code itself.
  The code change alone reaches an AUC of \AUCDiffOnly{}, and among commits of similar size the signal drops substantially, especially for \ClaudeCode.
  On the shared sample, \texttt{gpt-6-sol} reaches AUCs comparable to XGBoost and TF-IDF without any fine-tuning.

  \item \textbf{\rqc: \textit{\rqcc}}
  We hold out one agent at a time, together with the repositories in which it appears.
  Transfer is weakest for \ClaudeCode and \Copilot: XGBoost reaches AUCs of \AUCUnseenCC{} and \AUCUnseenCop{}, respectively, and TF-IDF performs worse for both.

  \item \textbf{\rqd: \textit{\rqdd}}
  We train on earlier months and test on later months of held-out repositories.
  \Copilot AUC falls from \AUCCopFirst{} to \AUCCopLow{} within \CopDropMonths{} months when fewer of its commits carry its squash-merge template.
  Training on more history helps \ClaudeCode but not \Copilot, and the TF-IDF signal fades unless the detector is retrained.
\end{itemize}

The results of our study show how incomplete attribution and limits in detection affect research datasets and development practice.
Researchers who build datasets from signatures should treat signature-based Agent labels as a lower bound and Human labels as the absence of recognized agent evidence, and they should compare Agent commits with Agent-era Human commits.
Researchers who evaluate authorship detectors should report precision at the observed prevalence and test on unseen agents and later months, because AUC alone overstates how useful a detector is.
Detector scores can help prioritize commits for review, but only disclosure identifies which commits an agent wrote.
To keep agent contributions traceable, coding agent builders should record attribution at commit time, since detection after the fact cannot recover removed signatures.



\section{Background}\label{sec:background}

\subsection{AI Coding Agents in Software Development}\label{subsec:coding_agents}

AI coding agents combine large language models with development tools to inspect repositories, edit files, run tests, and revise changes in response to feedback~\cite{yang2024sweagent,hassan2025agenticsoftwareengineering,anthropic2026claudecode,github2026copilotagent}.
This multi-step workflow extends beyond code completion, although developers can still guide and revise the agent's work.
\AIDev provides pull requests (PRs) labeled as agent-authored, together with repository metadata, reviews, and commits~\cite{li2026aidev}.
\tab{tab:agent_studies} summarizes empirical studies of coding agents on GitHub, including their topics, units of analysis, and methods for identifying agent-authored changes.



Coding agents are used in projects of different sizes, maturity levels, and programming languages, and agent-assisted commits are larger than commits classified as human-authored and often involve feature development and bug fixing~\cite{robbes2026agentic}.
PR acceptance rates vary by task type, with no single agent performing best on all tasks~\cite{pinna2026acceptance}, and agent-authored PRs that include tests are larger and take longer to complete than those without, at similar merge rates~\cite{haque2026tests}.
Task and change size therefore shape how agent and human contributions compare.
We account for them by comparing Agent and Human commits within the same repositories and era and by testing detection among commits of similar size.

\begin{table}[t]
  \centering
  \footnotesize
  \caption{Prior studies of AI coding agents and their methods for identifying agent contributions. The identification methods refer to dataset labels, and detector inputs may differ.}
  \label{tab:agent_studies}
  \begin{tabularx}{\linewidth}{lp{0.12\linewidth}X}
    \toprule
    Topics studied & Analysis unit & Agent identification \\
    \midrule
    Agent use and contribution characteristics~\cite{li2025aidev,robbes2026agentic}
      & Repositories, PRs, commits
      & Agent accounts, branch names, attribution messages, configuration files \\
    
    PR acceptance and developer intervention~\cite{DBLP:journals/corr/abs-2509-14745, pinna2026acceptance,khelifi2026interventions}
      & PRs
      & Attribution messages in PR descriptions or \AIDev labels \\

    Test inclusion, timing, and security~\cite{haque2026tests,siddiq2026security}
      & PRs, commits
      & \AIDev labels \\

    Code changes after merging~\cite{sawada2026maintenance}
      & Files, commits
      & \AIDev PRs and agent names in commit \\

    Agent attribution and detection~\cite{ghaleb2026fingerprinting,ghaleb2026agentag}
      & PRs
      & \AIDev labels \\
    \bottomrule
  \end{tabularx}
\end{table}


\subsection{Challenges in Attributing Agent Contributions}\label{subsec:agent_attribution}

\begin{figure}[b]
\centering
\includegraphics[width=0.6\linewidth]{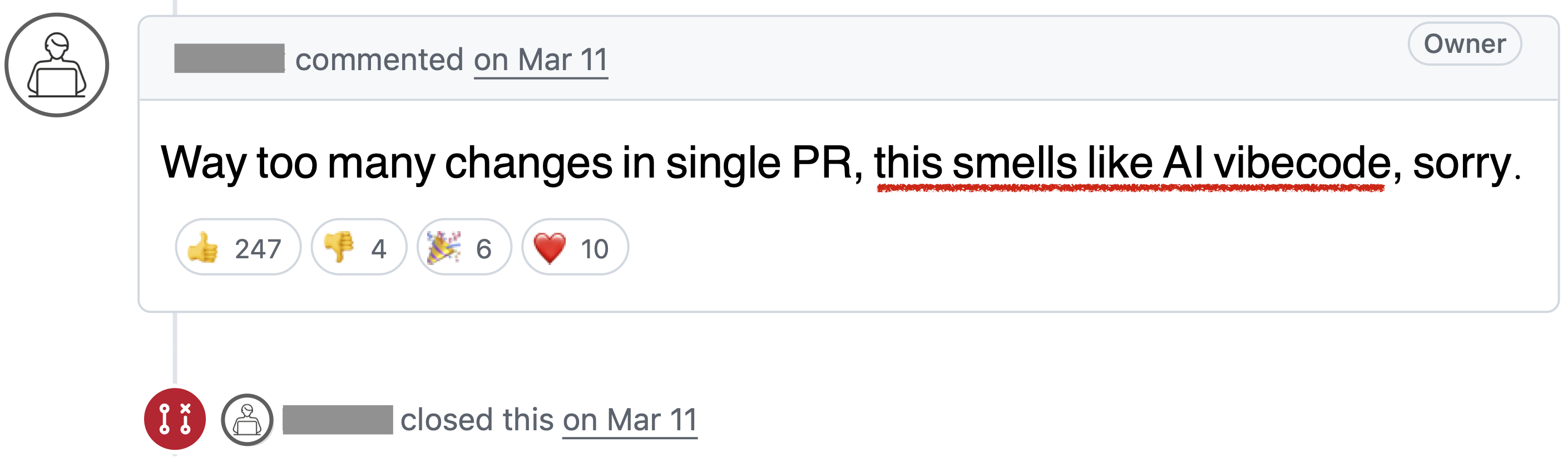}
\caption[Maintainer comment closing a suspected AI-generated PR]{Maintainer comment closing a PR suspected of being generated by an AI agent (\url{https://github.com/Fredolx/open-tv/pull/397}).}
\label{fig:motivating_example}
\Description{A maintainer closes a PR as suspected AI-generated work. The contributor reports using AI only for autocompletion; the exchange illustrates disagreement about authorship when commits lack agent signatures.}
\end{figure}

The PR in \fig{fig:motivating_example} illustrates the uncertainty in attributing contributions when agent signatures are absent.
The maintainer of this open-source media player closed the PR as likely AI-generated.
The PR changed \MotivFiles{} files and added \MotivLines{} lines, its commits carried no agent signature, and the contributor stated that AI tools had been used only for autocompletion.
Over \MotivEndorsements{} users endorsed the maintainer's comment.
A dataset that treats unsigned commits as Human would assign that label to these commits, but the available evidence does not resolve the differing accounts of how the code was produced.

This ambiguity reflects a limitation of signature-based attribution.
Previous studies use agent accounts, branch names, and attribution messages to identify agent-authored PRs and commits~\cite{li2025aidev,robbes2026agentic}.
However, developers can disable or remove attribution messages~\cite{robbes2026agentic}.
A PR submitted from a developer account can still contain agent-generated code.
In \AIDev, \AgenTagHumanMarkerPct{}\% of \NAgenTagHumanPRs{} PRs labeled human-authored contained explicit agent identifiers~\cite{ghaleb2026agentag}.
Neither a human account nor a missing agent identifier establishes that a code change was written without AI assistance.

Developers also revise agent-generated code during code review and subsequent maintenance.
In a study of \ClaudeCode PRs, \ClaudePRRevisedPct{}\% of merged PRs underwent further changes, including bug fixes and adjustments to project conventions~\cite{DBLP:journals/corr/abs-2509-14745}.
A study of agent-authored PRs found that guiding agents and enforcing project conventions were more common forms of human intervention than directly editing code~\cite{khelifi2026interventions}.
A study of software maintenance also found that human developers made most subsequent changes to agent-generated files~\cite{sawada2026maintenance}.
Consequently, labeling a PR as agent-authored does not establish who authored each commit in it or subsequent changes to the code.

\subsection{Detecting Agent-Authored Commits}\label{subsec:agent_detection}

Previous studies address two prediction tasks: identifying which agent authored a PR and separating agent-authored from human-authored PRs.
One study used PR-level features to identify the authoring agent among \NFingerprintAgents{} known agents~\cite{ghaleb2026fingerprinting}.
AgenTag combines PR text, commit messages, code changes, and behavioral features after removing direct agent identifiers~\cite{ghaleb2026agentag}.
In its balanced Agent--Human experiments, models using PR text and commit messages outperformed models using code diffs.
The PR-text model also separated the two classes when training and test repositories were disjoint, although this evaluation was limited to repositories containing both classes~\cite{ghaleb2026agentag}.
These findings motivate our joint analysis of commit-message and code-change features.
Detectors of LLM-generated code, such as GPTSniffer~\cite{nguyen2024gptsniffer} and DetectCodeGPT~\cite{shi2025between}, classify isolated snippets, whereas real agent commits mix generated, edited, and human-written changes.

Our study differs from this work in four ways:
(1) We classify individual commits rather than PRs, since AgenTag concatenates the commit messages of a PR and a PR label does not identify who wrote each commit~\cite{ghaleb2026agentag}.
(2) We evaluate detectors at the observed Agent share, where precision determines whether flagged commits are useful, whereas AgenTag reports balanced Agent--Human experiments.
(3) Our main protocol is repository-disjoint nested cross-validation, complemented by a month-by-month grid of training and test periods.
AgenTag evaluates its Agent--Human task with stratified cross-validation, checks repository separation only within repositories that contain both classes, and uses a chronological split only for agent identification.
(4) We test whether masking leaves merge-workflow artifacts that still identify agents.

\suppressfloats[t]
\begin{figure}[t]
\centering
\includegraphics[width=0.95\linewidth]{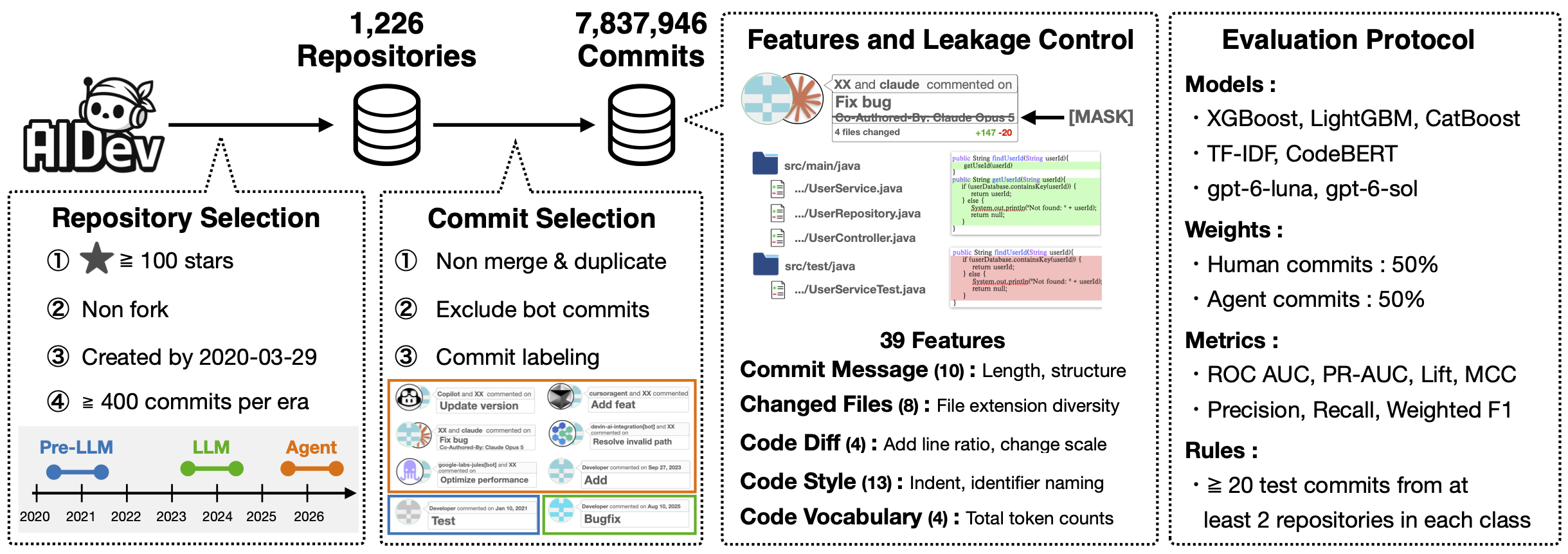}
\caption{Overview of the study design.}
\label{fig:overview}
\end{figure}

\section{Study Design}\label{sec:design}
This section describes the repository and commit selection, labeling, features, and evaluation protocol shared by all research questions (\fig{fig:overview}).

\subsection{Repository Selection}\label{subsec:repository_selection}

We identify \NCandidateRepos{} non-fork repositories in \AIDev \AIDevVersion{}~\cite{li2026aidev}, collected through \AIDevCutoff, with at least \MinStars{} stars and at least one PR by a studied agent.
We keep repositories created before \CreatedBefore{} (\NCreatedBeforeRepos{} candidates) with at least \MinCommitsPerEra{} default-branch commits in each \WindowMonths{}-month era (\tab{tab:dataset}), leaving \NScreenedRepos{} for collection.
Activity counts include all authors and merge commits.
The pre-LLM era ends at the technical preview of \Copilot on \CopilotPreviewLongDate{}~\cite{friedman2021copilot}.
The LLM era ends at the earliest PR by a studied agent in \AIDev, opened by \Devin on \FirstAgentPRDate{}~\cite{li2026aidev}.
The Agent era runs from \FirstMonth{}, the first full month after the public preview of the \CopilotFull on \CopilotAgentReleaseDate{}~\cite{github2025codingagentpreview} and the \Jules public beta on \JulesBetaDate{}~\cite{google2025jules}, to \LastMonth{}.


\subsection{Commit Selection}\label{subsec:commit_selection}

For each repository, we resolve its default branch (e.g., \texttt{main}) through the GitHub REST API~\cite{github2026restcommits}, record the head SHA of that branch and the retrieval time, and list the commits reachable from that SHA whose UTC committer dates fall within each era.
\AIDev supplies only the candidate repositories, and our own collection covers Agent and Human commits, including those after its cutoff.
We exclude merge commits with two or more parents, records missing required timestamps or parent information, and duplicate SHAs.
We apply no filter on squash merges or commit size.

We identify \Copilot, \Cursor, \Devin, and \Jules through the author accounts \texttt{Copilot}, \texttt{cursoragent}, \texttt{devin-ai-integration[bot]}, and \texttt{google-labs-jules[bot]}, respectively, and \ClaudeCode through a case-insensitive \texttt{Co-Authored-By: Claude} message signature.
These rules apply in every era.
We exclude \Codex because these rules do not identify its commits.
We exclude all copies of the \NMultiAgentCommits{} SHAs matching multiple agents.
Human commits carry no recognized AI-assistance signature and pass bot filtering based on Dey\etal~\cite{dey2020biman} and known automation accounts.
This label does not establish that no AI assistance was used.
The final dataset contains \NStudyCommits{} commits from \NCommonRepos{} repositories (\tab{tab:dataset}), because one selected repository retains no eligible commits.
Of these repositories, \NHumanEraRepos{} also have Human commits in both earlier eras.

\begin{table}[t]
  \centering
  \small

  \caption{Repositories, commits, and changes for the Agent and Human classes across the three eras: Pre-LLM (\PreWindowStart{} to \PreWindowLastDate{}), LLM (\LLMWindowStart{} to \LLMWindowLastDate{}), and Agent (\AgentWindowStart{} to \AgentWindowLastDate{}). Medians pool commits within each row; the repository-level medians in RQ1 give each repository equal weight.}
  \label{tab:dataset}
  \setlength{\tabcolsep}{4pt}
  \begin{tabular}{llrrrrrr}
    \toprule
    Era & Group & \#~Repos & \#~Commits & Median words & \#~Files & \#~LOC & Median LOC \\
    \midrule
    Pre-LLM & Human & \TabDatasetPreLLMHumanRepos{} & \TabDatasetPreLLMHumanCommits{} & \TabDatasetPreLLMHumanMedianWords{} & \TabDatasetPreLLMHumanFiles{} & \TabDatasetPreLLMHumanLOC{} & \TabDatasetPreLLMHumanMedianLOC{} \\
    \midrule
    LLM & Human & \TabDatasetLLMHumanRepos{} & \TabDatasetLLMHumanCommits{} & \TabDatasetLLMHumanMedianWords{} & \TabDatasetLLMHumanFiles{} & \TabDatasetLLMHumanLOC{} & \TabDatasetLLMHumanMedianLOC{} \\
    \midrule
    \multirow{6}{*}{Agent} & \Copilot & \TabDatasetAgentCopRepos{} & \TabDatasetAgentCopCommits{} & \TabDatasetAgentCopMedianWords{} & \TabDatasetAgentCopFiles{} & \TabDatasetAgentCopLOC{} & \TabDatasetAgentCopMedianLOC{} \\
    & \ClaudeCode & \TabDatasetAgentCCRepos{} & \TabDatasetAgentCCCommits{} & \TabDatasetAgentCCMedianWords{} & \TabDatasetAgentCCFiles{} & \TabDatasetAgentCCLOC{} & \TabDatasetAgentCCMedianLOC{} \\
    & \Cursor & \TabDatasetAgentCursorRepos{} & \TabDatasetAgentCursorCommits{} & \TabDatasetAgentCursorMedianWords{} & \TabDatasetAgentCursorFiles{} & \TabDatasetAgentCursorLOC{} & \TabDatasetAgentCursorMedianLOC{} \\
    & \Devin & \TabDatasetAgentDevinRepos{} & \TabDatasetAgentDevinCommits{} & \TabDatasetAgentDevinMedianWords{} & \TabDatasetAgentDevinFiles{} & \TabDatasetAgentDevinLOC{} & \TabDatasetAgentDevinMedianLOC{} \\
    & \Jules & \TabDatasetAgentJulesRepos{} & \TabDatasetAgentJulesCommits{} & \TabDatasetAgentJulesMedianWords{} & \TabDatasetAgentJulesFiles{} & \TabDatasetAgentJulesLOC{} & \TabDatasetAgentJulesMedianLOC{} \\
    & Human & \TabDatasetAgentHumanRepos{} & \TabDatasetAgentHumanCommits{} & \TabDatasetAgentHumanMedianWords{} & \TabDatasetAgentHumanFiles{} & \TabDatasetAgentHumanLOC{} & \TabDatasetAgentHumanMedianLOC{} \\
    \midrule
    Total & --- & \NCommonRepos{} & \TabDatasetTotalCommits{} & --- & \TabDatasetTotalFiles{} & \TabDatasetTotalLOC{} &  --- \\
    \bottomrule
  \end{tabular}
\end{table}

\subsection{Features and Leakage Control}\label{subsec:features}

\tab{tab:features} summarizes the \NFeatures{} detector features: \NPriorFeatures{} adapt prior-work measures to individual commits and \NNewFeatures{} use exploratory definitions.
Commit Message features follow prior work on commit message quality and conventions~\cite{tian2022commit}.
Full definitions and provenance accompany the replication package~\cite{artifact_agent_or_human}.
For \rqa{}, we also summarize changed files, added/deleted/total changed lines, and three timing measures (hour, weekday, weekend).
Dates and timing are not detector inputs.

\suppressfloats[t]
\begin{table}[t]
  \centering
  \small
  \caption{Overview of detector features.}
  \label{tab:features}
  \begin{tabular}{l l r }
    \toprule
    Feature family & Features & Count \\
    \midrule
    Commit Message & Length~\cite{ghaleb2026agentag}, formatting, Markdown structure, and writing conventions~\cite{conventionalcommits} & \NMsgFeatures \\
    Changed Files & Extension diversity~\cite{dey2020biman} and file-type fractions & \NFileFeatures \\
    Code Diff & Addition fraction, changes per file, single-file and large-change indicators & \NDiffFeatures \\
    Code Style & Added-line count~\cite{shi2025between}, layout~\cite{caliskan2015stylometry}, identifiers, and annotation proxies & \NStyleFeatures \\
    Code Vocabulary & Token counts, Zipf and Heaps exponents~\cite{shi2025between} & \NLexFeatures \\
    \bottomrule
  \end{tabular}
\end{table}

The first three feature families use commit metadata.
Code Style and Code Vocabulary need the full patch, so we compute them only for patch-eligible commits, whose patches cover every changed file and line and match the recorded line counts.
Before feature extraction, we remove lines containing agent attributions, generation notices, tool URLs, or vendor templates, then replace remaining tool names with a placeholder~\cite{ghaleb2026agentag}.
Unmasked inputs provide a sensitivity check.

Within each training fold, we drop constant or entirely missing features, impute remaining missing values with training medians, and apply AutoSpearman~\cite{jiarpakdee2018autospearman} to remove features with high correlation ($|\rho|\geq\CorrThreshold{}$) or multicollinearity (VIF $\geq\VIFThreshold{}$), so that SHAP values and ablation effects are not split across correlated features.
The fitted preprocessing is then applied unchanged to the validation and test data.
For comparison, we also evaluate the unreduced feature set.

\subsection{Evaluation Protocol}\label{subsec:evaluation}

Our main detector is XGBoost~\cite{DBLP:conf/kdd/ChenG16} on the selected numeric features.
Human commits receive half of the training weight, and the other half is divided equally among the agents, so no commit is discarded.
We compare it with LightGBM~\cite{ke2017lightgbm} and CatBoost~\cite{prokhorenkova2018catboost} on the same features and with TF-IDF term weighting~\cite{salton1988tfidf} over hashed tokens~\cite{weinberger2009hashing} of the masked message and diff, of the message alone, and of the diff alone, classified by $\ell_2$-regularized logistic regression with the same weights.
The TF-IDF input also excludes file paths, the mask token, \texttt{Co-Authored-By} trailers, and ``Initial plan'' lines, which act as residual signatures.
We also ask LLM judges~\cite{zheng2023judging}, without fine-tuning, to score the masked message and diff of a stratified sample of \NSampleAgent{} Agent and \NSampleHuman{} Human commits, and we evaluate XGBoost and TF-IDF on the same sample.
We query \texttt{gpt-6-luna}~\cite{openai2026gpt6luna} zero-shot and \texttt{gpt-6-sol}~\cite{openai2026gpt6sol} zero-shot and few-shot with \NFewShot{} fixed labeled examples from repositories outside the sample, and remove \texttt{Co-Authored-By} trailers and ``Initial plan'' lines from the \texttt{gpt-6-sol} prompts.
The prompts are in the directory \texttt{prompts/llm\_judge} of our replication package~\cite{artifact_agent_or_human}.

The LLM judges use a fixed decision threshold of \DefaultThreshold{}.
Their precision is re-weighted from the sample's true- and false-positive rates to the observed Agent share (\AgentShare{}).
On the same sample, we evaluate logistic regression on frozen CodeBERT~\cite{DBLP:conf/emnlp/FengGTDFGS0LJZ20} embeddings of the masked message and diff, of the message alone, and of the diff alone, trained on \NCodeBERTTrain{} commits from the training repositories of each outer fold.

\textit{Splits and thresholds.}
Training and test repositories are disjoint.
\rqb{} uses \NFolds{} outer folds.
\rqc{} additionally holds out one agent, and \rqd{} keeps the repository folds fixed and trains on months before those it tests.
Hyperparameters maximize mean per-agent AUC on repository-disjoint inner validation data.
Learned detectors use the threshold that maximizes F1 on validation data and fix it before testing.
\rqc{} and \rqd{} restrict their validation data further.

\textit{Measures and uncertainty.}
Agent is the positive class.
We report ROC AUC and the area under the precision--recall curve (PR-AUC, computed as non-interpolated average precision), together with precision, recall, and F1 at each test set's observed Agent share, because ROC AUC alone can overstate performance on imbalanced data~\cite{davis2006relationship,saito2015precision}.
Unlike ROC AUC, PR-AUC depends on the Agent share, and a random detector reaches a PR-AUC equal to that share.
To compare agents with different shares, we therefore also report lift, the ratio of PR-AUC to the Agent share.
We also report the Matthews correlation coefficient (MCC), which uses all four cells of the confusion matrix, and weighted F1, the measure reported by AgenTag~\cite{ghaleb2026agentag}.
Results cover all agents pooled and each agent against the same Human commits.
An agent is evaluable only with at least \MinEvalCommits{} test commits from at least \MinEvalRepos{} repositories in each class.
We use \NBootstrap{} bootstrap samples~\cite{efron1979bootstrap} that resample whole repositories~\cite{field2007cluster} for 95\% confidence intervals and we compare detectors on same test commits using the same bootstrap samples for both.

\section{Results}\label{sec:results}

For each research question we present the motivation, the approach, and the results.


\subsection{\rqa: \rqaa}\label{sec:rq_characteristics}

\subsubsection{Motivation.}
To detect Agent commits from their content, we first need to understand how they differ from Human commits.
The differences may lie in the code, in the commit message, or in the conventions of a particular tool.
Each has a different consequence for detection.
A detector that relies on the code would keep working across tools, while one that relies on message templates would break when a tool changes its template.
The choice of Human reference also matters.
Human commits may have changed since LLM assistants became available, so a comparison with older Human commits could blame agents for a broader shift in practice.
This RQ aims to study which features separate Agent commits from Human commits of the same repositories and era, and whether Human commits themselves drifted over the same period.

\subsubsection{Approach.}
We summarize each feature within each repository and group using medians for continuous features and proportions for binary features.
For each agent, we keep repositories with at least \MinRQOneCommits{} commits from that agent, \MinRQOneCommits{} Agent-era Human commits, and \MinRQOneCommits{} pre-LLM-era Human commits.
We retain repositories with all three feature summaries and report median paired differences and median within-repository Cliff's $\delta$~\cite{Cliff,Cliff_threshold}, where positive values indicate higher Agent values.
We interpret $|\delta|$ below \CliffNegligible{}, \CliffSmall{}, and \CliffMedium{} as negligible, small, and medium, and larger values as large~\cite{Cliff_threshold}.
For the Human-only comparison, we use repositories with observed summaries in all three eras.
Each repository receives equal weight.

\rqa{} uses \NBootstrapRQOne{} repository-bootstrap draws for pointwise 95\% intervals and two-sided Wilcoxon signed-rank tests~\cite{Wilcoxon1945} of paired summaries when at least \MinWilcoxonRepos{} repositories are available.
Benjamini--Hochberg adjustment~\cite{benjamini1995controlling} treats the Human-era comparisons and the Agent--Human comparisons of all features as two separate families.
\tab{tab:rq1_cliffs_delta} reports features with $|\delta|\geq\DeltaDisplayCutoff{}$ for at least one of \ClaudeCode, \Copilot, and \Cursor (\NReposRQOneCC{}, \NReposRQOneCop{}, and \NReposRQOneCursor{} matched repositories), retaining all three estimates in each selected row.
This is a display cutoff, not a significance criterion.
\Devin and \Jules have only one and four matched repositories, so their results, like the confidence intervals and detailed tests for all agents, are in the supplementary materials.

\subsubsection{Results.}

\finding{Message-length features show the largest Agent--Human differences.}
\tab{tab:rq1_cliffs_delta} shows that character, word, and line counts have the largest effects for all three agents, and the word-count difference is significant for each of them.
For \ClaudeCode, the median over repositories of each repository's median message length is \MsgWordsCC{} words, compared with \MsgWordsHuman{} for Agent-era Human commits, so a typical \ClaudeCode message is about seven times as long as a typical Human message in the same repository.
The difference lies in length rather than format, e.g., the share of messages with a conventional prefix differs little between groups with negligible effect size ($|\delta|$ below \CliffNegligible{}) for all three agents.

\begin{table}[t]
  \small
  \centering
  \caption{Cliff's $\delta$ between Agent and Agent-era Human commits, as the median over repositories. Positive values mean higher values for Agent commits, and $^{*}$ marks an adjusted $p<0.05$. Only features with $|\delta|\geq \DeltaDisplayCutoff{}$ for at least one agent are shown. Orange marks positive and blue negative effects.}
  \label{tab:rq1_cliffs_delta}
\begin{tabular}{llrrr}
\toprule
Family & Feature & Claude Code & Copilot & Cursor \\
\midrule
\multirow{4}{*}{Commit Message} & Characters & \cellcolor[HTML]{EFCDA4}$\TabEffectsCharactersCC{}$\rlap{$^{*}$} & \cellcolor[HTML]{F3DABD}$\TabEffectsCharactersCop{}$\rlap{$^{*}$} & \cellcolor[HTML]{EDC89C}$\TabEffectsCharactersCursor{}$\rlap{$^{*}$} \\
& Words & \cellcolor[HTML]{EFCDA5}$\TabEffectsWordsCC{}$\rlap{$^{*}$} & \cellcolor[HTML]{F5E1C8}$\TabEffectsWordsCop{}$\rlap{$^{*}$} & \cellcolor[HTML]{EFCDA5}$\TabEffectsWordsCursor{}$\rlap{$^{*}$} \\
& Lines & \cellcolor[HTML]{F0D1AC}$\TabEffectsLinesCC{}$\rlap{$^{*}$} & \cellcolor[HTML]{F2D8B8}$\TabEffectsLinesCop{}$\rlap{$^{*}$} & \cellcolor[HTML]{EDC699}$\TabEffectsLinesCursor{}$\rlap{$^{*}$} \\
& Conventional prefix & \cellcolor[HTML]{FFFEFC}$\TabEffectsConventionalPrefixCC{}$\rlap{$^{*}$} & \cellcolor[HTML]{FFFFFF}$\TabEffectsConventionalPrefixCop{}$ & \cellcolor[HTML]{FCF5EC}$\TabEffectsConventionalPrefixCursor{}$ \\
\addlinespace[3pt]
\multirow{3}{*}{Code Diff} & Addition share & \cellcolor[HTML]{FCF4EC}$\TabEffectsAdditionShareCC{}$\rlap{$^{*}$} & \cellcolor[HTML]{FCF6EF}$\TabEffectsAdditionShareCop{}$\rlap{$^{*}$} & \cellcolor[HTML]{FBF4EB}$\TabEffectsAdditionShareCursor{}$ \\
& Changed lines per file & \cellcolor[HTML]{FBF1E6}$\TabEffectsChangedLinesPerFileCC{}$\rlap{$^{*}$} & \cellcolor[HTML]{FEFAF7}$\TabEffectsChangedLinesPerFileCop{}$\rlap{$^{*}$} & \cellcolor[HTML]{FDF9F3}$\TabEffectsChangedLinesPerFileCursor{}$ \\
& Single-file commit & \cellcolor[HTML]{F7FAFC}$\TabEffectsSingleFileCommitCC{}$\rlap{$^{*}$} & \cellcolor[HTML]{F6F8FC}$\TabEffectsSingleFileCommitCop{}$\rlap{$^{*}$} & \cellcolor[HTML]{FFFEFD}$\TabEffectsSingleFileCommitCursor{}$ \\
\addlinespace[3pt]
\multirow{4}{*}{Code Style} & Added lines in patches & \cellcolor[HTML]{FAEFE3}$\TabEffectsAddedLinesInPatchesCC{}$\rlap{$^{*}$} & \cellcolor[HTML]{FDF7F1}$\TabEffectsAddedLinesInPatchesCop{}$\rlap{$^{*}$} & \cellcolor[HTML]{FDF8F3}$\TabEffectsAddedLinesInPatchesCursor{}$ \\
& Identifiers & \cellcolor[HTML]{F9EEE0}$\TabEffectsIdentifiersCC{}$\rlap{$^{*}$} & \cellcolor[HTML]{FCF6EF}$\TabEffectsIdentifiersCop{}$\rlap{$^{*}$} & \cellcolor[HTML]{FDF8F2}$\TabEffectsIdentifiersCursor{}$ \\
& camelCase share & \cellcolor[HTML]{FEFCFA}$\TabEffectsCamelCaseShareCC{}$\rlap{$^{*}$} & \cellcolor[HTML]{FFFEFE}$\TabEffectsCamelCaseShareCop{}$ & \cellcolor[HTML]{FCF7F1}$\TabEffectsCamelCaseShareCursor{}$ \\
& Indentation variance & \cellcolor[HTML]{FCF7F0}$\TabEffectsIndentationVarianceCC{}$\rlap{$^{*}$} & \cellcolor[HTML]{FEFCF9}$\TabEffectsIndentationVarianceCop{}$\rlap{$^{*}$} & \cellcolor[HTML]{FEFDFB}$\TabEffectsIndentationVarianceCursor{}$ \\
\addlinespace[3pt]
\multirow{2}{*}{Code Vocabulary} & Tokens & \cellcolor[HTML]{FAEEE0}$\TabEffectsTokensCC{}$\rlap{$^{*}$} & \cellcolor[HTML]{FCF7F1}$\TabEffectsTokensCop{}$\rlap{$^{*}$} & \cellcolor[HTML]{FDF8F3}$\TabEffectsTokensCursor{}$ \\
& Distinct tokens & \cellcolor[HTML]{F9ECDD}$\TabEffectsDistinctTokensCC{}$\rlap{$^{*}$} & \cellcolor[HTML]{FDF8F2}$\TabEffectsDistinctTokensCop{}$\rlap{$^{*}$} & \cellcolor[HTML]{FDF9F4}$\TabEffectsDistinctTokensCursor{}$ \\
\bottomrule
\end{tabular}

\end{table}

\finding{Differences in the code change are smaller and vary by agent.}
\ClaudeCode and \Copilot have more distinct added tokens and more changed lines per file than Human commits, whereas the \Cursor intervals for these features span zero.
These effects are small or negligible, whereas the word-count effects are medium to large.
No Changed Files feature reaches the display cutoff for any of the three agents.
These observations describe the labeled commits, without establishing code quality or detection performance.

\finding{Human commits also changed across eras.}
\tab{tab:rq1_human_eras} shows that Human messages became longer and Human changes touched more lines per file from the pre-LLM to the Agent era across \NHumanEraRepos{} repositories.
Commit Message features already rose from the pre-LLM to the LLM era, whereas Code Style and Code Diff measures such as added lines and changed lines per file shifted mainly from the LLM to the Agent era.
A pre-LLM-era Human reference would therefore mix era changes with Agent--Human differences, which supports comparing Agent commits with Agent-era Human commits.

\begin{table}[t]
  \small
  \centering
  \caption{Changes in Human commits between eras in the same \NHumanEraRepos{} repositories, as matched-pairs rank-biserial correlations~\cite{kerby2014rankbiserial}. Positive values mean higher values in the later era, and $^{*}$ marks an adjusted $p<0.05$. Rows are the features of \tab{tab:rq1_cliffs_delta}. The correlation measures how consistently repositories change in one direction, not the size of the change.}
  \label{tab:rq1_human_eras}
\begin{tabular}{llrrr}
\toprule
Family & Feature & pre-LLM to LLM & LLM to Agent & pre-LLM to Agent \\
\midrule
\multirow{4}{*}{Commit Message} & Characters & \cellcolor[HTML]{F6E2CB}$\TabErasCharactersPreToLLM{}$\rlap{$^{*}$} & \cellcolor[HTML]{F2D6B6}$\TabErasCharactersLLMToAgent{}$\rlap{$^{*}$} & \cellcolor[HTML]{F0CFA9}$\TabErasCharactersPreToAgent{}$\rlap{$^{*}$} \\
& Words & \cellcolor[HTML]{F6E4CF}$\TabErasWordsPreToLLM{}$\rlap{$^{*}$} & \cellcolor[HTML]{F2D6B5}$\TabErasWordsLLMToAgent{}$\rlap{$^{*}$} & \cellcolor[HTML]{F1D2AD}$\TabErasWordsPreToAgent{}$\rlap{$^{*}$} \\
& Lines & \cellcolor[HTML]{F3DABD}$\TabErasLinesPreToLLM{}$\rlap{$^{*}$} & \cellcolor[HTML]{F2D6B4}$\TabErasLinesLLMToAgent{}$\rlap{$^{*}$} & \cellcolor[HTML]{EFCDA5}$\TabErasLinesPreToAgent{}$\rlap{$^{*}$} \\
& Conventional prefix & \cellcolor[HTML]{F2D6B5}$\TabErasConventionalPrefixPreToLLM{}$\rlap{$^{*}$} & \cellcolor[HTML]{EFCEA7}$\TabErasConventionalPrefixLLMToAgent{}$\rlap{$^{*}$} & \cellcolor[HTML]{ECC393}$\TabErasConventionalPrefixPreToAgent{}$\rlap{$^{*}$} \\
\addlinespace[3pt]
\multirow{3}{*}{Code Diff} & Addition share & \cellcolor[HTML]{FFFEFE}$\TabErasAdditionSharePreToLLM{}$ & \cellcolor[HTML]{F0D1AD}$\TabErasAdditionShareLLMToAgent{}$\rlap{$^{*}$} & \cellcolor[HTML]{F3D9BB}$\TabErasAdditionSharePreToAgent{}$\rlap{$^{*}$} \\
& Changed lines per file & \cellcolor[HTML]{FEFDFB}$\TabErasChangedLinesPerFilePreToLLM{}$ & \cellcolor[HTML]{EECA9F}$\TabErasChangedLinesPerFileLLMToAgent{}$\rlap{$^{*}$} & \cellcolor[HTML]{F1D2AD}$\TabErasChangedLinesPerFilePreToAgent{}$\rlap{$^{*}$} \\
& Single-file commit & \cellcolor[HTML]{EAF0F8}$\TabErasSingleFileCommitPreToLLM{}$\rlap{$^{*}$} & \cellcolor[HTML]{D9E4F2}$\TabErasSingleFileCommitLLMToAgent{}$\rlap{$^{*}$} & \cellcolor[HTML]{D3E0F0}$\TabErasSingleFileCommitPreToAgent{}$\rlap{$^{*}$} \\
\addlinespace[3pt]
\multirow{4}{*}{Code Style} & Added lines in patches & \cellcolor[HTML]{FCF6EF}$\TabErasAddedLinesInPatchesPreToLLM{}$\rlap{$^{*}$} & \cellcolor[HTML]{EFCDA5}$\TabErasAddedLinesInPatchesLLMToAgent{}$\rlap{$^{*}$} & \cellcolor[HTML]{F1D2AD}$\TabErasAddedLinesInPatchesPreToAgent{}$\rlap{$^{*}$} \\
& Identifiers & \cellcolor[HTML]{FCF6EF}$\TabErasIdentifiersPreToLLM{}$\rlap{$^{*}$} & \cellcolor[HTML]{EFCCA4}$\TabErasIdentifiersLLMToAgent{}$\rlap{$^{*}$} & \cellcolor[HTML]{F0D0AA}$\TabErasIdentifiersPreToAgent{}$\rlap{$^{*}$} \\
& camelCase share & \cellcolor[HTML]{FDF9F4}$\TabErasCamelCaseSharePreToLLM{}$\rlap{$^{*}$} & \cellcolor[HTML]{FDF8F2}$\TabErasCamelCaseShareLLMToAgent{}$\rlap{$^{*}$} & \cellcolor[HTML]{FCF6F0}$\TabErasCamelCaseSharePreToAgent{}$\rlap{$^{*}$} \\
& Indentation variance & \cellcolor[HTML]{FEFDFB}$\TabErasIndentationVariancePreToLLM{}$ & \cellcolor[HTML]{F6E2CA}$\TabErasIndentationVarianceLLMToAgent{}$\rlap{$^{*}$} & \cellcolor[HTML]{F8E8D6}$\TabErasIndentationVariancePreToAgent{}$\rlap{$^{*}$} \\
\addlinespace[3pt]
\multirow{2}{*}{Code Vocabulary} & Tokens & \cellcolor[HTML]{FCF5EE}$\TabErasTokensPreToLLM{}$\rlap{$^{*}$} & \cellcolor[HTML]{EFCDA4}$\TabErasTokensLLMToAgent{}$\rlap{$^{*}$} & \cellcolor[HTML]{F0D0A9}$\TabErasTokensPreToAgent{}$\rlap{$^{*}$} \\
& Distinct tokens & \cellcolor[HTML]{FDF8F1}$\TabErasDistinctTokensPreToLLM{}$\rlap{$^{*}$} & \cellcolor[HTML]{EECA9F}$\TabErasDistinctTokensLLMToAgent{}$\rlap{$^{*}$} & \cellcolor[HTML]{F0CFA8}$\TabErasDistinctTokensPreToAgent{}$\rlap{$^{*}$} \\
\bottomrule
\end{tabular}

\end{table}

\summarybox{Answer to \rqa{}}{Message-length features show the largest differences between the three displayed agents and Agent-era Human commits. Code-change differences are smaller and vary by agent. Human commits also change across eras, making Agent-era Human commits the primary reference for detection.}

\subsection{\rqb: \rqbb}\label{sec:rq_detection}

\subsubsection{Motivation.}
Signatures identify Agent commits only when an agent leaves one, otherwise, the commit content is the only available evidence.
\rqa{} showed that Agent and Human commits differ as groups, but a group difference does not guarantee that individual commits can be classified, especially in repositories the detector has never seen.
The goal of this RQ is to understand how far a content-based detector can be trusted on commits without a signature, and which part of its evidence would survive a change in message conventions.

\subsubsection{Approach.}
We train a binary detector to distinguish Agent commits from Agent-era Human commits.
Commits from all five studied agents share the positive Agent label, Human is the negative label, and the detector does not predict which agent authored a commit.
The main detector uses all five feature families on patch-eligible commits, with masked inputs and AutoSpearman.
\rqc{} and \rqd{} use the same masking and correlation reduction.
To measure what Code Style and Code Vocabulary add, we also train a metadata-only model on the Commit Message, Changed Files, and Code Diff features and compare the two models on the same patch-eligible commits with paired repository bootstrapping.

We compare masked and unmasked inputs on the same commits and evaluate both full and correlation-reduced feature sets.
To examine feature contributions, we compute SHAP values~\cite{lundberg2017shap} for held-out XGBoost predictions and remove one feature family at a time.
Each ablation repeats feature selection, hyperparameter tuning, and model fitting within the training folds.
These ablations use masked inputs and the correlation-reduced model.

\textit{Size-matched evaluation.}
We evaluate the main detector without retraining on commits of similar size.
We restrict its out-of-fold predictions to held-out Agent--Human pairs matched by the same repository, language, and month, and by predefined size ranges.
The ranges are \SizeBinsWords{} message words, \SizeBinsLines{} changed lines, and \SizeBinsFiles{} changed files.
Sizes are matched by range, not exactly.

\textit{Baselines and sensitivity checks.}
We fit a baseline using only message word count, changed lines, and changed files.
LightGBM, CatBoost, and TF-IDF use the same commits and folds as XGBoost, while the LLM judge uses the sample described in \sec{subsec:evaluation}.
We also repeat the evaluation with Agent commits weighted equally, keeping the total weight of the Agent and Human classes equal.

\textit{Review budgets.}
We use review fractions of \BudgetSmall{}\%, \BudgetMid{}\%, and \BudgetLarge{}\% per test fold.
Within each fold, we select the highest scores, round counts up, and resolve ties by SHA order.
Precision and recall are calculated from the combined counts across folds.


\subsubsection{Results.}

\begin{table}[t]
  \centering
  \footnotesize
  \caption{Detector performance on masked commits, using the full test set (\NRQTwoTest{} commits) or a balanced sample (\NSample{} commits). Detectors within each set use the same commits. Lift is PR-AUC divided by the Agent share. Weighted F1 averages Agent and Human F1 by class size. The sample's PR-AUC, lift, and weighted F1 are not comparable with those of the full test set. Its precision and MCC are re-weighted to the observed Agent share of \AgentShare{}. CodeBERT is trained on \NCodeBERTTrain{} commits per fold.
  }
  \label{tab:rq2_models}
  \setlength{\tabcolsep}{3.5pt}
  \begin{tabular}{llrrrrrrr}
    \toprule
    Test set & Detector & ROC AUC & PR-AUC & Lift & Precision & Recall & MCC & Weighted F1 \\
    \midrule
    \multirow{6}{*}{Full} & XGBoost, \NReducedFeatures{} features & \TabModelsFullXGBAUC{} & \TabModelsFullXGBPRAUC{} & \TabModelsFullXGBLift{} & \TabModelsFullXGBPrec{} & \TabModelsFullXGBRec{} & \TabModelsFullXGBMCC{} & \TabModelsFullXGBWFOne{} \\
    & LightGBM, \NReducedFeatures{} features & \TabModelsFullLGBMAUC{} & \TabModelsFullLGBMPRAUC{} & \TabModelsFullLGBMLift{} & \TabModelsFullLGBMPrec{} & \TabModelsFullLGBMRec{} & \TabModelsFullLGBMMCC{} & \TabModelsFullLGBMWFOne{} \\
    & CatBoost, \NReducedFeatures{} features & \TabModelsFullCatBoostAUC{} & \TabModelsFullCatBoostPRAUC{} & \TabModelsFullCatBoostLift{} & \TabModelsFullCatBoostPrec{} & \TabModelsFullCatBoostRec{} & \TabModelsFullCatBoostMCC{} & \TabModelsFullCatBoostWFOne{} \\ 
    & TF-IDF, message and diff & \textbf{\TabModelsFullTFIDFAUC{}} & \TabModelsFullTFIDFPRAUC{} & \TabModelsFullTFIDFLift{} & \textbf{\TabModelsFullTFIDFPrec{}} & \TabModelsFullTFIDFRec{} & \TabModelsFullTFIDFMCC{} & \textbf{\TabModelsFullTFIDFWFOne{}} \\
    & TF-IDF, message only & \textbf{\TabModelsFullTFIDFMsgAUC{}} & \textbf{\TabModelsFullTFIDFMsgPRAUC{}} & \textbf{\TabModelsFullTFIDFMsgLift{}} & \TabModelsFullTFIDFMsgPrec{} & \textbf{\TabModelsFullTFIDFMsgRec{}} & \textbf{\TabModelsFullTFIDFMsgMCC{}} & \TabModelsFullTFIDFMsgWFOne{} \\
    & TF-IDF, diff only & \TabModelsFullTFIDFDiffAUC{} & \TabModelsFullTFIDFDiffPRAUC{} & \TabModelsFullTFIDFDiffLift{} & \TabModelsFullTFIDFDiffPrec{} & \TabModelsFullTFIDFDiffRec{} & \TabModelsFullTFIDFDiffMCC{} & \TabModelsFullTFIDFDiffWFOne{} \\
    \midrule
    \multirow{12}{*}{Sample} & XGBoost, \NReducedFeatures{} features & \TabModelsSampleXGBAUC{} & \TabModelsSampleXGBPRAUC{} & \TabModelsSampleXGBLift{} & \TabModelsSampleXGBPrec{} & \TabModelsSampleXGBRec{} & \TabModelsSampleXGBMCC{} & \TabModelsSampleXGBWFOne{} \\
    & LightGBM, \NReducedFeatures{} features & \TabModelsSampleLGBMAUC{} & \TabModelsSampleLGBMPRAUC{} & \TabModelsSampleLGBMLift{} & \TabModelsSampleLGBMPrec{} & \TabModelsSampleLGBMRec{} & \TabModelsSampleLGBMMCC{} & \TabModelsSampleLGBMWFOne{} \\
    & CatBoost, \NReducedFeatures{} features & \TabModelsSampleCatBoostAUC{} & \TabModelsSampleCatBoostPRAUC{} & \TabModelsSampleCatBoostLift{} & \TabModelsSampleCatBoostPrec{} & \TabModelsSampleCatBoostRec{} & \TabModelsSampleCatBoostMCC{} & \TabModelsSampleCatBoostWFOne{} \\
    & TF-IDF, message and diff & \TabModelsSampleTFIDFAUC{} & \TabModelsSampleTFIDFPRAUC{} & \TabModelsSampleTFIDFLift{} & \textbf{\TabModelsSampleTFIDFPrec{}} & \TabModelsSampleTFIDFRec{} & \TabModelsSampleTFIDFMCC{} & \TabModelsSampleTFIDFWFOne{} \\
    & TF-IDF, message only & \TabModelsSampleTFIDFMsgAUC{} & \TabModelsSampleTFIDFMsgPRAUC{} & \TabModelsSampleTFIDFMsgLift{} & \TabModelsSampleTFIDFMsgPrec{} & \TabModelsSampleTFIDFMsgRec{} & \TabModelsSampleTFIDFMsgMCC{} & \TabModelsSampleTFIDFMsgWFOne{} \\
    & TF-IDF, diff only & \TabModelsSampleTFIDFDiffAUC{} & \TabModelsSampleTFIDFDiffPRAUC{} & \TabModelsSampleTFIDFDiffLift{} & \TabModelsSampleTFIDFDiffPrec{} & \TabModelsSampleTFIDFDiffRec{} & \TabModelsSampleTFIDFDiffMCC{} & \TabModelsSampleTFIDFDiffWFOne{} \\
    & CodeBERT, message and diff & \TabModelsSampleCodeBERTAUC{} & \TabModelsSampleCodeBERTPRAUC{} & \TabModelsSampleCodeBERTLift{} & \TabModelsSampleCodeBERTPrec{} & \TabModelsSampleCodeBERTRec{} & \TabModelsSampleCodeBERTMCC{} & \TabModelsSampleCodeBERTWFOne{} \\
    & CodeBERT, message only & \textbf{\TabModelsSampleCodeBERTMsgAUC{}} & \textbf{\TabModelsSampleCodeBERTMsgPRAUC{}} & \textbf{\TabModelsSampleCodeBERTMsgLift{}} & \TabModelsSampleCodeBERTMsgPrec{} & \TabModelsSampleCodeBERTMsgRec{} & \textbf{\TabModelsSampleCodeBERTMsgMCC{}} & \textbf{\TabModelsSampleCodeBERTMsgWFOne{}} \\
    & CodeBERT, diff only & \TabModelsSampleCodeBERTDiffAUC{} & \TabModelsSampleCodeBERTDiffPRAUC{} & \TabModelsSampleCodeBERTDiffLift{} & \TabModelsSampleCodeBERTDiffPrec{} & \textbf{\TabModelsSampleCodeBERTDiffRec{}} & \TabModelsSampleCodeBERTDiffMCC{} & \TabModelsSampleCodeBERTDiffWFOne{} \\
    & LLM zero-shot (gpt-6-luna) & \TabModelsSampleLunaZeroAUC{} & \TabModelsSampleLunaZeroPRAUC{} & \TabModelsSampleLunaZeroLift{} & \TabModelsSampleLunaZeroPrec{} & \TabModelsSampleLunaZeroRec{} & \TabModelsSampleLunaZeroMCC{} & \TabModelsSampleLunaZeroWFOne{} \\
    & LLM zero-shot (gpt-6-sol) & \TabModelsSampleSolZeroAUC{} & \TabModelsSampleSolZeroPRAUC{} & \TabModelsSampleSolZeroLift{} & \TabModelsSampleSolZeroPrec{} & \TabModelsSampleSolZeroRec{} & \TabModelsSampleSolZeroMCC{} & \TabModelsSampleSolZeroWFOne{} \\
    & LLM few-shot (gpt-6-sol) & \TabModelsSampleSolFewAUC{} & \TabModelsSampleSolFewPRAUC{} & \TabModelsSampleSolFewLift{} & \TabModelsSampleSolFewPrec{} & \TabModelsSampleSolFewRec{} & \TabModelsSampleSolFewMCC{} & \TabModelsSampleSolFewWFOne{} \\
    \bottomrule
  \end{tabular}
\end{table}


\finding{Masked detectors rank Agent commits above Human commits, but even the best reaches \PrecTFIDF{} precision.}
\tab{tab:rq2_models} compares the detectors on the same \NRQTwoTest{} commits, with an Agent share of \AgentShare{}.
XGBoost on the \NReducedFeatures{} features reaches ROC AUC \AUCXGB{}, and TF-IDF on masked messages and diffs reaches \AUCTFIDF{}.
At validation-selected thresholds, however, their precision is only \PrecXGB{} and \PrecTFIDF{}, so most flagged commits are Human commits.
Per agent, AUC ranges from \AUCXGBCC{} to \AUCAgentMax{} and lift from \LiftAgentMin{} to \LiftAgentMax{}, yet precision falls from \PrecXGBCC{} for \ClaudeCode to \PrecSmallAgentsMax{} or less for \Cursor, \Devin, and \Jules, whose Agent shares are below \ShareSmallAgentsMax{} (\tab{tab:rq2_agents}).
Even the top \BudgetSmall{}\% of XGBoost scores in each fold reaches a precision of only \TopOnePctPrec{}, with larger budgets giving lower precision.
Weighted F1, the measure reported by AgenTag~\cite{ghaleb2026agentag}, remains high in both settings.
On the balanced sample, it reaches up to \WFOneCodeBERTMessage{}, close to the \WFOneAgenTag{} that AgenTag reports for commit messages on balanced PRs.
On the full test set, TF-IDF reaches \WFOneTFIDFFull{} because the Human class dominates the average, although its precision is only \PrecTFIDF{}.

Without fine-tuning, \texttt{gpt-6-sol} reaches AUCs similar to XGBoost and TF-IDF, and CodeBERT performs best on the sample when its input includes the message.
On the sample of \NSample{} commits, \texttt{gpt-6-sol} with \NFewShot{} examples reaches AUC \AUCSolFew{}, compared with \AUCXGBSample{} for XGBoost, and the paired interval of the difference includes zero.
CodeBERT embeddings of the message and diff reach AUC \AUCCodeBERT{}, above XGBoost with a paired interval that excludes zero, but only \AUCCodeBERTCode{} on the diff alone.
The learned representation thus mirrors the message--code split of the handcrafted features.

\finding{Other classifiers and training weights reach similar AUCs, and correlation reduction lowers AUC.}
LightGBM and CatBoost reach AUCs of \AUCLightGBM{} and \AUCCatBoost{}, and equal weights for all Agent commits give \AUCEqualWeight{}. 
Without correlation reduction, XGBoost reaches \AUCNoAutoSpearman{}, so the main model trades some AUC for interpretable feature contributions.
Signatures left in the input raise these values.
Without masking, XGBoost reaches \UnmaskedAUC{}, and TF-IDF reaches \AUCTFIDFAllLines{} when file paths, the mask token, and workflow lines are kept.
A detector trained against LLM-era Human commits reaches \AUCLLMEraHuman{}, consistent with the changes in Human commits across eras in \rqa{}.

\begin{figure}[!h]
  \centering
  \includegraphics[width=0.7\linewidth]{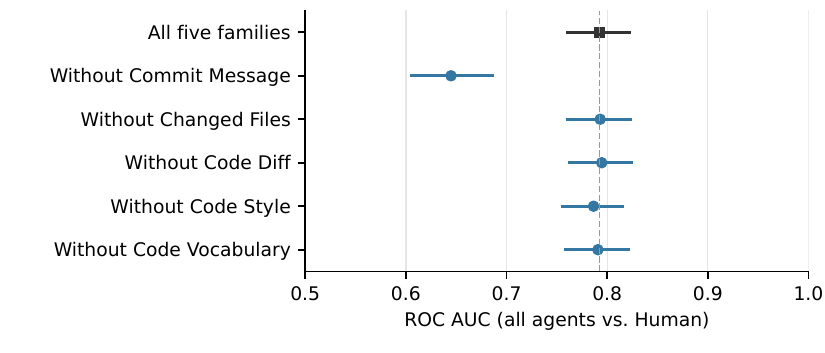}
  \caption{XGBoost ROC AUC when one feature family is removed, with 95\% repository-bootstrap intervals. The dashed line marks the model with all five families.}
  \label{fig:rq2_ablations}
  \Description{Removing Commit Message features lowers overall binary ROC AUC from \AUCXGB{} to \AUCNoMessage{}. Other family removals leave AUC between \AUCAblationMin{} and \AUCAblationMax{}.}
\end{figure}

\finding{The signal lies mainly in the commit message, with change size as the second source.}
Without the message, detection weakens sharply.
\tab{tab:rq2_models} shows that TF-IDF on the diff alone reaches AUC \AUCDiffOnly{} and CodeBERT on the diff alone reaches \AUCCodeBERTCode{}, so the learned representation shows the same split between message and code as the handcrafted features.
\fig{fig:rq2_ablations} shows that removing the Commit Message features lowers XGBoost from \AUCXGB{} to \AUCNoMessage{}, whereas removing any other feature family barely changes it. Message length in lines has the largest held-out SHAP importance in every outer fold, and adding Code Style and Code Vocabulary features raises AUC by only \PatchGain{} on the same commits.

\begin{table}[!t]
  \centering
  \small
  \caption{XGBoost detection of each agent against the same \NRQTwoHuman{} Human commits. Lift is PR-AUC divided by the Agent share. The last column is ROC AUC on size-matched test pairs, which pair each Agent commit with a Human commit from the same repository, language, month, and size range.}
  \label{tab:rq2_agents}
  \setlength{\tabcolsep}{4pt}
  \begin{tabular}{lrrrrrrrr}
    \toprule
    & \multicolumn{7}{c}{Full test set} & Size-matched \\
    \cmidrule(lr){2-8}\cmidrule(l){9-9}
    Agent subset & Agent share & ROC AUC & PR-AUC & Lift & Precision & Recall & MCC & ROC AUC \\
    \midrule
    \ClaudeCode & \TabAgentsCCShare{} & \TabAgentsCCAUC{} & \TabAgentsCCPRAUC{} & \TabAgentsCCLift{} & \TabAgentsCCPrec{} & \TabAgentsCCRec{} & \TabAgentsCCMCC{} & \TabAgentsCCMatchedAUC{} \\
    \Copilot & \TabAgentsCopShare{} & \TabAgentsCopAUC{} & \TabAgentsCopPRAUC{} & \TabAgentsCopLift{} & \TabAgentsCopPrec{} & \TabAgentsCopRec{} & \TabAgentsCopMCC{} & \TabAgentsCopMatchedAUC{} \\
    \Cursor & \TabAgentsCursorShare{} & \TabAgentsCursorAUC{} & \TabAgentsCursorPRAUC{} & \TabAgentsCursorLift{} & \TabAgentsCursorPrec{} & \TabAgentsCursorRec{} & \TabAgentsCursorMCC{} & \TabAgentsCursorMatchedAUC{} \\
    \Devin & \TabAgentsDevinShare{} & \TabAgentsDevinAUC{} & \TabAgentsDevinPRAUC{} & \TabAgentsDevinLift{} & \TabAgentsDevinPrec{} & \TabAgentsDevinRec{} & \TabAgentsDevinMCC{} & \TabAgentsDevinMatchedAUC{} \\
    \Jules & \TabAgentsJulesShare{} & \TabAgentsJulesAUC{} & \TabAgentsJulesPRAUC{} & \TabAgentsJulesLift{} & \TabAgentsJulesPrec{} & \TabAgentsJulesRec{} & \TabAgentsJulesMCC{} & \TabAgentsJulesMatchedAUC{} \\
    \bottomrule
  \end{tabular}
\end{table}

\finding{Among commits of similar size, both detectors lose most of their signal.} 
\tab{tab:rq2_agents} shows that on test pairs matched by repository, language, month, and size range, \ClaudeCode AUC falls from \AUCXGBCC{} to \AUCMatchedCCXGB{} for XGBoost and from \AUCTFIDFCC{} to \AUCMatchedCCTFIDF{} for TF-IDF.
\Copilot keeps more of its signal, with \AUCMatchedCopXGB{} and \AUCMatchedCopTFIDF{}, and TF-IDF on the diff alone reaches only \AUCMatchedCCDiff{} for \ClaudeCode and \AUCMatchedCopDiff{} for \Copilot, close to chance.
A baseline using only message word count, changed lines, and changed files reaches AUC \AUCSizeOnly{} on the full test set, and for \ClaudeCode the full XGBoost detector does not outperform it.
Most of the signal therefore comes from message length and change size, consistent with the drop without Commit Message features in \fig{fig:rq2_ablations}.

Merge-workflow lines inflate this result if they are left in the message.
When \texttt{Co-Authored-By} trailers and \Copilot's ``Initial plan'' lines are kept, TF-IDF reaches \AUCMatchedCopWorkflow{} for \Copilot on the matched pairs.
The share of sampled \Copilot messages with a \texttt{Co-Authored-By} trailer is \CoauthorCopShare{}, compared with \CoauthorHumanShare{} for Human messages, so these lines act as residual signatures after masking.

\summarybox{Answer to \rqb{}}{After signatures are masked, detectability comes mostly from message verbosity and change size. The code change alone reaches an AUC of \AUCDiffOnly{}, and among commits of similar size the signal drops substantially, especially for \ClaudeCode. Detectors still rank Agent commits above Human commits with an AUC of up to \AUCTFIDF{}, and \texttt{gpt-6-sol} without fine-tuning reaches AUCs similar to XGBoost and TF-IDF on the shared sample. Even the best detector on the full test set reaches a precision of only \PrecTFIDF{} at the observed Agent share of \AgentShare{}, so detector scores support prioritizing commits for review rather than labeling them.}

\subsection{\rqc: \rqcc}\label{sec:rq_generalization}

\subsubsection{Motivation.}
A deployed detector will encounter agents it was not trained on, including agents using new tools and tools whose commits carry no signature that we could label.
Good performance on the training agents does not tell us whether the detector learned characteristics shared by all agents or exclusively the habits of the agents it observed.
If it learned only those habits, applying it broadly could miss unrepresented agents and skew the resulting datasets.
This \rqc{} aims to understand how well a detector ranks commits from an agent it has never seen.

\subsubsection{Approach.}
We use the masked XGBoost model with correlation reduction in leave-one-agent-out evaluation.
This evaluation uses all eligible commits from the Agent era, including those without complete patches.
For each held-out agent, the test set contains its commits and Human commits from repositories in which that agent appears.
We exclude all commits from these repositories from training, regardless of their labels.
The remaining Human commits and commits from the other four agents form the training set.

Each row in \tab{tab:unseen_agent_performance} represents a separate experiment.
The training Agent count covers the other four agents, whereas the test Agent count covers only the agent named in that row.
For example, the \ClaudeCode row trains on \NMinAgentTrain{} commits from other agents and tests on \NClaudeCode{} \ClaudeCode commits, alongside the Human commits in the respective columns.
Training counts vary because each held-out agent appears in a different set of repositories, all of which must be excluded from its training set.

\begin{table}[t]
  \centering
  \caption{Commits in each unseen-agent experiment. Training and test repositories are disjoint.}
  \label{tab:unseen_agent_performance}

  \setlength{\tabcolsep}{4pt}
  \begin{tabular}{lrrrr}
    \toprule
    Unseen agent & \shortstack{Training\\Other agents} & \shortstack{Training\\Human} & \shortstack{Test\\Unseen agent} & \shortstack{Test\\Human} \\

    \midrule
    \ClaudeCode & \TabTrainingCCTrainAgent{} & \TabTrainingCCTrainHuman{} & \TabTrainingCCTestAgent{} & \TabTrainingCCTestHuman{} \\
    \Copilot & \TabTrainingCopTrainAgent{} & \TabTrainingCopTrainHuman{} & \TabTrainingCopTestAgent{} & \TabTrainingCopTestHuman{} \\
    \Cursor & \TabTrainingCursorTrainAgent{} & \TabTrainingCursorTrainHuman{} & \TabTrainingCursorTestAgent{} & \TabTrainingCursorTestHuman{} \\
    \Devin & \TabTrainingDevinTrainAgent{} & \TabTrainingDevinTrainHuman{} & \TabTrainingDevinTestAgent{} & \TabTrainingDevinTestHuman{} \\
    \Jules & \TabTrainingJulesTrainAgent{} & \TabTrainingJulesTrainHuman{} & \TabTrainingJulesTestAgent{} & \TabTrainingJulesTestHuman{} \\
    \bottomrule
  \end{tabular}
\end{table}

Hyperparameters and the decision threshold are selected on repository-disjoint validation within this training set, so they are chosen using known agents only.
The tuning objective balances the agents present in each training fold.
Boxplots summarize the repository-bootstrap AUC draws of each fitted model, not independent model fits.
We also report a diagnostic at a threshold of \DefaultThreshold{}.

To check whether differences in training volume explain performance across agents, we repeat the evaluation with \NMinAgentTrain{} Agent and \NMinHumanTrain{} Human training commits, the smallest available class-specific training pools.
We retain each training agent and allocate the Agent budget proportionally to its available commits.
We use three sampling seeds and report their median AUC, keeping the test commits and the full model's hyperparameters unchanged.

We perform two additional sensitivity checks.
One allows training commits from other agents in the test repositories, while still excluding the held-out agent and all test commits, and tunes these models separately.
The other restricts training to patch-eligible commits and compares both models on common patch-eligible test commits under the original repository split.

\subsubsection{Results.}

\begin{figure}[!h]
  \centering
  \includegraphics[width=0.72\linewidth]{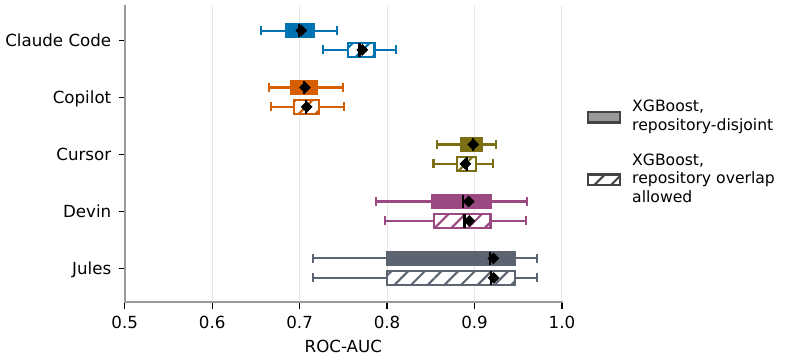}
  \caption{ROC AUC on unseen agents. Boxes and whiskers summarize repository-bootstrap draws (median, interquartile range, and 95\% interval), and diamonds show the test-set AUC.}
  \label{fig:unseen_agent_auc}
  \Description{Two horizontal boxplots per unseen agent show repository-bootstrap AUC draws for XGBoost: repository-disjoint training (filled) and repository overlap allowed (hatched). These distributions condition on fixed fitted models.}
\end{figure}

\finding{For the two largest agents, detection drops to AUC \AUCUnseenCC{} and \AUCUnseenCop{} when the agent is absent from training.}
All confidence intervals in \fig{fig:unseen_agent_auc} lie above \RandomAUC{}, so each held-out agent is still ranked above Human commits.
The gap is not a training-volume effect, because the AUCs remain almost unchanged with training counts fixed.
\Cursor, \Devin, and \Jules reach AUCs of \AUCUnseenSmallMin{} or higher, but these estimates rest on few test commits and repositories, and they may partly reflect the verbosity of \ClaudeCode commits in their training data rather than a signal specific to those agents.

\finding{The vocabulary signal transfers weakly to held-out \ClaudeCode and \Copilot.}
\tab{tab:rq3_tfidf} shows that the best detector~(i.e., TF-IDF) on the full test set in \rqb{}, ranks unseen-agent commits worse than XGBoost on the same test commits.
Its AUC is only \AUCUnseenCCTFIDF{} for \ClaudeCode and \AUCUnseenCopTFIDF{} for \Copilot, and its paired intervals lie below XGBoost for these two agents and for \Cursor.
For these two agents, the vocabulary signal is weak, while the \NReducedFeatures{} features, which rely mostly on message length and change size, keep more of their ranking.

\begin{table}[t]
  \centering
  \small
  \caption{Detection of unseen agents, with both detectors on the same test commits and thresholds selected on known agents.}
  \label{tab:rq3_tfidf}
  \setlength{\tabcolsep}{4pt}
  \begin{tabular}{llrrrrrrr}
    \toprule
    Unseen agent & Agent share & Detector & ROC AUC & PR-AUC & Lift & Precision & Recall & MCC \\
    \midrule
    \multirow{2}{*}{\ClaudeCode} & \multirow{2}{*}{\TabUnseenCCShare{}} & XGBoost & \TabUnseenCCXGBAUC{} & \TabUnseenCCXGBPRAUC{} & \TabUnseenCCXGBLift{} & \TabUnseenCCXGBPrec{} & \TabUnseenCCXGBRec{} & \TabUnseenCCXGBMCC{} \\
     & & TF-IDF & \TabUnseenCCTFIDFAUC{} & \TabUnseenCCTFIDFPRAUC{} & \TabUnseenCCTFIDFLift{} & \TabUnseenCCTFIDFPrec{} & \TabUnseenCCTFIDFRec{} & \TabUnseenCCTFIDFMCC{} \\
    \midrule
    \multirow{2}{*}{\Copilot} & \multirow{2}{*}{\TabUnseenCopShare{}} & XGBoost & \TabUnseenCopXGBAUC{} & \TabUnseenCopXGBPRAUC{} & \TabUnseenCopXGBLift{} & \TabUnseenCopXGBPrec{} & \TabUnseenCopXGBRec{} & \TabUnseenCopXGBMCC{} \\
     & & TF-IDF & \TabUnseenCopTFIDFAUC{} & \TabUnseenCopTFIDFPRAUC{} & \TabUnseenCopTFIDFLift{} & \TabUnseenCopTFIDFPrec{} & \TabUnseenCopTFIDFRec{} & \TabUnseenCopTFIDFMCC{} \\
    \midrule
    \multirow{2}{*}{\Cursor} & \multirow{2}{*}{\TabUnseenCursorShare{}} & XGBoost & \TabUnseenCursorXGBAUC{} & \TabUnseenCursorXGBPRAUC{} & \TabUnseenCursorXGBLift{} & \TabUnseenCursorXGBPrec{} & \TabUnseenCursorXGBRec{} & \TabUnseenCursorXGBMCC{} \\
     & & TF-IDF & \TabUnseenCursorTFIDFAUC{} & \TabUnseenCursorTFIDFPRAUC{} & \TabUnseenCursorTFIDFLift{} & \TabUnseenCursorTFIDFPrec{} & \TabUnseenCursorTFIDFRec{} & \TabUnseenCursorTFIDFMCC{} \\
    \midrule
    \multirow{2}{*}{\Devin} & \multirow{2}{*}{\TabUnseenDevinShare{}} & XGBoost & \TabUnseenDevinXGBAUC{} & \TabUnseenDevinXGBPRAUC{} & \TabUnseenDevinXGBLift{} & \TabUnseenDevinXGBPrec{} & \TabUnseenDevinXGBRec{} & \TabUnseenDevinXGBMCC{} \\
     & & TF-IDF & \TabUnseenDevinTFIDFAUC{} & \TabUnseenDevinTFIDFPRAUC{} & \TabUnseenDevinTFIDFLift{} & \TabUnseenDevinTFIDFPrec{} & \TabUnseenDevinTFIDFRec{} & \TabUnseenDevinTFIDFMCC{} \\
    \midrule
    \multirow{2}{*}{\Jules} & \multirow{2}{*}{\TabUnseenJulesShare{}} & XGBoost & \TabUnseenJulesXGBAUC{} & \TabUnseenJulesXGBPRAUC{} & \TabUnseenJulesXGBLift{} & \TabUnseenJulesXGBPrec{} & \TabUnseenJulesXGBRec{} & \TabUnseenJulesXGBMCC{} \\
     & & TF-IDF & \TabUnseenJulesTFIDFAUC{} & \TabUnseenJulesTFIDFPRAUC{} & \TabUnseenJulesTFIDFLift{} & \TabUnseenJulesTFIDFPrec{} & \TabUnseenJulesTFIDFRec{} & \TabUnseenJulesTFIDFMCC{} \\
    \bottomrule
  \end{tabular}
\end{table}

\finding{Thresholds selected on known agents transfer poorly to unseen agents.}
\tab{tab:rq3_tfidf} also reports performance at the threshold selected on known agents.
For \ClaudeCode, this threshold gives a recall of only \RecUnseenCC{}, compared with \RecUnseenCCDefault{} at a threshold of \DefaultThreshold{}.
For \Jules, an AUC of \AUCUnseenJules{} comes with a precision of \PrecUnseenJules{}, so most flagged commits are Human commits. The gap is not a training-volume effect, because the AUCs remain almost unchanged with training counts fixed at \NMinAgentTrain{} Agent and \NMinHumanTrain{} Human commits.
Allowing other-agent training commits from the test repositories raises \ClaudeCode AUC to \AUCUnseenCCRelaxed{} and barely changes the other agents.
Restricting training to patch-eligible commits also barely changes AUC for every agent.



\summarybox{Answer to \rqc{}}{Detection transfers poorly to held-out \ClaudeCode and \Copilot, with AUCs of \AUCUnseenCC{} and \AUCUnseenCop{}, whereas \Cursor, \Devin, and \Jules reach XGBoost AUCs of at least \AUCUnseenSmallMin{} on smaller test sets. TF-IDF falls further for \ClaudeCode and \Copilot, showing weak vocabulary transfer. XGBoost's threshold tuned on known agents recalls only \RecUnseenCC{} of unseen \ClaudeCode commits, with precision below \PrecUnseenMax{} for every held-out agent.}

\subsection{\rqd: \rqdd}\label{sec:rq_temporal}

\subsubsection{Motivation.}
A detector used in practice must classify commits written after its training data were collected. For example, agents may change their behavior and message templates and repositories may change their workflows. If detector performance drifts over time, an estimate of agent activity based on the detector could rise or fall even when the actual level of activity remains unchanged. This RQ aims to measure how detector performance changes in the months following training and whether a longer training history slows this change.

\subsubsection{Approach.}
We split the Agent era into its \WindowMonths{} calendar months.
\rqd{} uses the patch-eligible commits and the main detector of \rqb{}.
Because the relation between features and labels can change over time~\cite{gama2014drift}, we use an expanding training window, a strategy studied in long-lived defect prediction~\cite{mcintosh2018moving}.
For each monthly cutoff, we train on all commits from the start of the era through that month, excluding the held-out repositories, then freeze the model and its preprocessing and evaluate each later month on commits from the held-out repositories.
For example, a model trained on June is evaluated on July and later months, while a model trained on June and July is evaluated on August and later months.
We do not evaluate the training month or earlier months.

Each repository is assigned to one of \NFolds{} outer folds for the whole era.
For each outer fold, hyperparameters are selected on the training repositories' first-month commits and the decision threshold on inner validation in \FirstMonth{}, and both stay fixed for all later training cutoffs and test months.
Threshold results therefore show how a threshold chosen once transfers to later months.
For each training window and test month, we pool out-of-fold predictions across folds.
We report \ClaudeCode and \Copilot, the only agents meeting the support rule in every training-test month pair.
Each test month uses the same commits for every training cutoff.

We present a heatmap with cumulative training windows as rows and future test months as columns.
Each row shows the future performance of a fixed model, while each column compares models trained on different amounts of history against the same test commits.
For the final test month, \LastMonth{}, we compare the earliest and latest training cutoffs on identical test commits using \NBootstrap{} paired repository-bootstrap draws, and we also report PR-AUC, precision, recall, and F1 for these two training histories.

\subsubsection{Results.}

\begin{figure}[t]
  \centering
  \includegraphics[width=\linewidth]{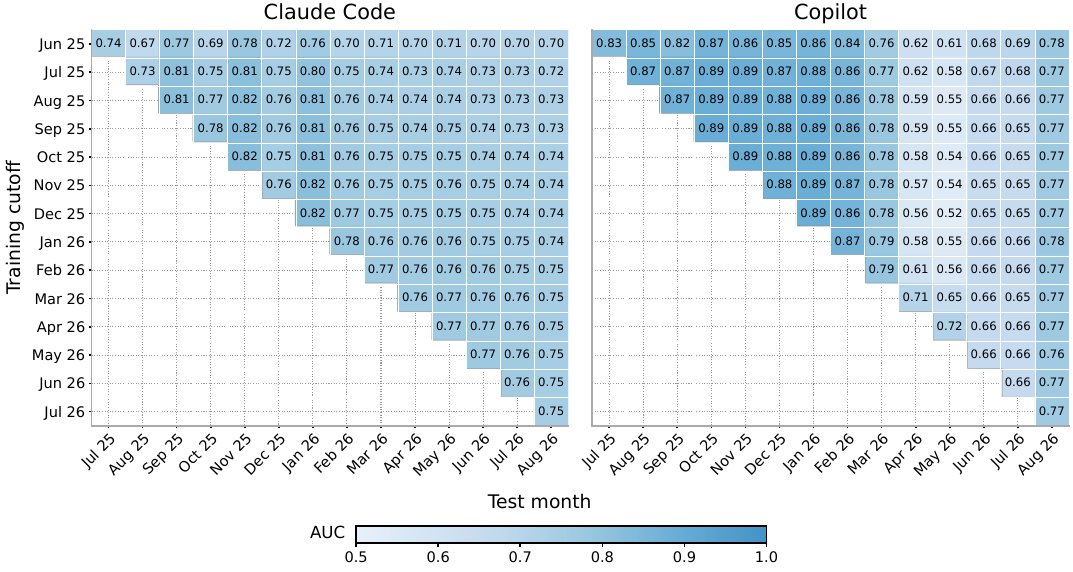}
  \caption{ROC AUC of \ClaudeCode and \Copilot by training cutoff (rows) and test month (columns). White cells are not evaluated.}
  \label{fig:temporal_auc}
  \Description{Two heatmaps show Claude Code and Copilot AUCs for all \NMonthPairs{} pairs of training cutoff and later test month. Each cell prints its AUC.}
\end{figure}

\finding{Detection performance falls and recovers across test periods, even without retraining.}
\fig{fig:temporal_auc} shows all \NMonthPairs{} training-test month pairs, and its first row keeps training fixed to \FirstMonth{}.
For this frozen model, \Copilot AUC falls from \AUCCopFirst{} in \FirstTestMonth{} to \AUCCopLow{} in \CopLowMonth{}, then recovers to \AUCCopLast{} in \LastMonth{} without any update to the model.
\ClaudeCode AUC stays between \AUCCCFrozenMin{} and \AUCCCFrozenMax{}.
The \Copilot dip also appears in the same test months across training cutoffs, including models trained on more recent commits.
These patterns suggest that the performance changes come from differences between test periods rather than from elapsed time since training alone.
One possible explanation for this dip is a change in \Copilot's workflow templates.
During the period of lower AUC, fewer \Copilot commits carried the squash-merge template of its PR workflow.
The share of \Copilot messages with an ``Initial plan'' line fell from \InitialPlanShareHigh{} in \InitialPlanHighMonth{} to \InitialPlanShareLow{} in \InitialPlanLowMonth{} and then recovered, and the share with squash-merge bullet lists followed the same path.
\begin{table}[t]
  \centering
  \caption{Detection in \LastMonth{} by training cutoff. Training starts in \FirstMonth{}. Test commits: \NRQFourFinalCC{} \ClaudeCode (Agent share \ShareCCLast{}) and \NRQFourFinalCop{} \Copilot (\ShareCopLast{}), each against the same \NRQFourFinalHuman{} Human commits.}
  \label{tab:rq4_final_month}
  \begin{tabular}{llrrrrr}
    \toprule
    Agent & Training cutoff & ROC AUC & PR-AUC & Precision & Recall & F1 \\
    \midrule
    \multirow{2}{*}{\ClaudeCode} & \FirstMonth{} & \TabFinalCCFirstAUC{} & \TabFinalCCFirstPRAUC{} & \TabFinalCCFirstPrec{} & \TabFinalCCFirstRec{} & \TabFinalCCFirstFOne{} \\
     & \LastTrainMonth{} & \TabFinalCCLastAUC{} & \TabFinalCCLastPRAUC{} & \TabFinalCCLastPrec{} & \TabFinalCCLastRec{} & \TabFinalCCLastFOne{} \\
    \midrule
    \multirow{2}{*}{\Copilot} & \FirstMonth{} & \TabFinalCopFirstAUC{} & \TabFinalCopFirstPRAUC{} & \TabFinalCopFirstPrec{} & \TabFinalCopFirstRec{} & \TabFinalCopFirstFOne{} \\
     & \LastTrainMonth{} & \TabFinalCopLastAUC{} & \TabFinalCopLastPRAUC{} & \TabFinalCopLastPrec{} & \TabFinalCopLastRec{} & \TabFinalCopLastFOne{} \\
    \bottomrule
  \end{tabular}
\end{table}

\finding{Expanding the training history improves \ClaudeCode detection but yields no clear AUC gain for \Copilot.}
\tab{tab:rq4_final_month} shows that on the same \LastMonth{} commits, training through \LastTrainMonth{} instead of \FirstMonth{} alone raises \ClaudeCode AUC from \AUCCCFinalFrozen{} to \AUCCCFinalRetrained{}, with a paired interval that excludes zero.
Recall more than doubles at a similar precision.
For \Copilot, the paired interval of the AUC difference includes zero, and its precision falls.

We notice that updating the training data helps the \ClaudeCode vocabulary detector more.
For \ClaudeCode, TF-IDF trained on \FirstMonth{} falls from AUC \AUCTFIDFFrozenFirst{} in \FirstTestMonth{} to \AUCTFIDFFrozenLast{} in \LastMonth{}, and it is below XGBoost in \NMonthsTFIDFBelow{} of the \NTestMonths{} later months.
Trained through \LastTrainMonth{}, it instead reaches \AUCTFIDFRetrained{}, above XGBoost with a paired interval that excludes zero.
For \Copilot, TF-IDF has a lower AUC than XGBoost with either training history, although the paired intervals of these differences include zero.
Expanding the training history can therefore recover detection performance, most clearly for \ClaudeCode's vocabulary signal, but the gain depends on the agent and the detector.

\finding{Above-chance ranking persists, but a fixed threshold misses most \ClaudeCode commits.}
With the model and thresholds fixed in \FirstMonth{}, \ClaudeCode AUC remains between \AUCCCFrozenMin{} and \AUCCCFrozenMax{}, yet recall never exceeds \RecCCFrozenMax{} in any later month.
Its precision rises from \PrecCCFrozenFirst{} to \PrecCCFrozenLast{} as the \ClaudeCode share of test commits grows from \ShareCCFirst{} to \ShareCCLast{}.
Higher prevalence can increase precision without improving ranking, so this rise in precision does not by itself show better detection.
For \Copilot, precision falls after \CopShiftMonth{} together with its AUC.
\tab{tab:rq4_final_month} shows that even with the expanded training history, recall stays low for both agents at the original thresholds, and most flagged commits are Human commits.
These results show why later-month evaluations need precision and recall alongside AUC, especially when detector predictions are used to estimate agent activity.


\summarybox{Answer to \rqd{}}{Detector performance is not stable over time. \Copilot AUC falls and recovers even without retraining, with the dip coinciding with fewer workflow templates in its messages. Expanding the training history improves \ClaudeCode detection, especially with TF-IDF, but yields no clear XGBoost AUC gain for \Copilot.}

\section{Implications}\label{sec:discussion}

\subsection{Implications for Researchers}

\textbf{Detectors capture how specific agents describe changes, not what they change.}
The code change alone gives an AUC of \AUCDiffOnly{} with TF-IDF and \AUCCodeBERTCode{} with CodeBERT embeddings, and among commits of similar size its signal falls close to chance, whereas the commit message carries most of the signal (\rqb{}).
This signal is also specific to agents and time.
TF-IDF separates known agents with an AUC of \AUCTFIDF{} but reaches only \AUCUnseenCCTFIDF{} for an unseen \ClaudeCode (\rqc{}), and its AUC falls from \AUCTFIDFFrozenFirst{} to \AUCTFIDFFrozenLast{} unless it is retrained (\rqd{}).
Because the code representations we evaluated separate agent and human changes only weakly, attribution research should study how changes are described, and evaluations should report unseen-agent and later-month results.

\smallskip\noindent\textbf{Detection performance depends on the evaluation setting.}
Class-weighted measures can look strong while precision stays low.
Weighted F1 reaches up to \WFOneCodeBERTMessage{} on a balanced sample and \WFOneTFIDFFull{} on the full test set, whereas at the validation-selected thresholds precision at the observed Agent share is at most \PrecTFIDF{} (\rqb{}).
Keeping \texttt{Co-Authored-By} trailers and ``Initial plan'' lines in the input raises the TF-IDF AUC for \Copilot on size-matched pairs from \AUCMatchedCopTFIDF{} to \AUCMatchedCopWorkflow{} (\rqb{}).
Even the strongest detector on the \NSample{}-commit sample, CodeBERT on the message alone, reaches an AUC of \AUCCodeBERTMessage{} with a precision of only \PrecCodeBERT{} at the observed Agent share, so the detectors we evaluated do not substitute for missing signatures.
Detection studies can complement balanced results with precision and PR-AUC at the observed Agent share and with masking of workflow artifacts, and studies built on signature-based labels can treat Agent labels as a lower bound.

\subsection{Implications for Developers}

\textbf{Detector scores can prioritize review but cannot establish authorship.}
Reviewing the highest-scoring \BudgetSmall{}\% of commits gives a precision of \TopOnePctPrec{} at a recall of \TopOnePctRec{}, and even the best detector on the full test set reaches a precision of \PrecTFIDF{} at the observed Agent share (\rqb{}).
Maintainers who need to know which commits an agent wrote should require disclosure, for example a trailer checked in continuous integration, rather than infer it afterward.

\smallskip\noindent\textbf{Human commits are moving toward agent-like commits.}
Human commits grew longer between the pre-LLM and Agent eras (\rqa{}), and detection against Agent-era Human commits is harder than against LLM-era Human commits (\rqb{}).
If this trend reduces the message-length contrast, post-hoc detection may become harder, so projects that rely on attribution for licensing, security review, or contribution credit need it recorded at commit time.

\subsection{Implications for Coding Agent Builders}

\textbf{Optional attribution is hard to recover once it is removed.}
Agent signatures are tool settings that users can change or disable, and the detectors we evaluated do not close the resulting gap from commit content.
Agent tools that aim to keep contributions traceable should record attribution in a form that survives edits and squash-merges, such as signed commit metadata, rather than in an optional message trailer.


\smallskip\noindent\textbf{Message verbosity is a visible and changeable fingerprint.}
The repository-level median \ClaudeCode commit message has \MsgWordsCC{} words, compared with \MsgWordsHuman{} for Agent-era Human commits (\rqa{}), and message length drives most detection.
Builders should be aware that their default message style identifies their tool today, and that changes to it can make earlier detectors obsolete.

\section{Threats to Validity}\label{sec:threats}

\subsection{Construct Validity}
Agent labels identify recognized accounts or authorship signatures, while Human denotes their absence after bot filtering.
Neither establishes exclusive authorship, because undisclosed AI use and squashed Human--Agent contributions can cross these boundaries.
The direction of resulting performance bias is unknown.
Features also reflect tasks and repository conventions, so differences characterize the labeled groups rather than authorship alone.
Detecting commits whose signatures were masked is not the same as detecting undisclosed agent use.
Historical Human commits may also include AI-assisted changes.
The LLM judges may have seen some public commits during pretraining, which could inflate their scores, although they did not exceed the trained detectors.

\subsection{Internal Validity}
Comparisons are observational, for example, eras differ in contributors and practices, size matching changes the test population, and training histories change volume, recency, and agent composition together.
Equal training counts do not equalize tasks or repositories.
Separately tuned models also differ in preprocessing, training populations, and thresholds.
Performance depends on these choices, including correlation reduction.
Feature ablations and SHAP values describe predictions, not causal effects.
Additional metrics, detectors, historical comparisons, count controls, paired-test families, and review-budget analyses were specified after initial results and were not preregistered.
Bootstrap intervals hold the fitted models fixed, so they exclude the uncertainty of retraining and tuning.
Repeated samples that share test commits are not independent replications.
Intervals are pointwise, and intervals spanning zero do not establish equivalence.
Signed-rank tests and false-discovery correction also rely on symmetry and dependence assumptions.

\subsection{External Validity}
Selection from \AIDev favors mature, active public repositories and does not represent private or recently created projects.
Collecting from the recorded default-branch head misses commits that were earlier removed from its history, and retrieval failures and incomplete patches can further affect which commits enter each evaluation.
\Devin and \Jules have few labeled commits, limiting the stability of their estimates.
Only \ClaudeCode and \Copilot have enough commits in every month for the temporal analysis.
Transfer conclusions therefore concern the evaluated repositories, agents, periods, and label definitions.

\section{Conclusion}\label{sec:conclusion}

Empirical studies of AI coding agents identify agent commits through signatures that developers can remove, and it has been unclear whether agent commits can be recognized without them.
We studied \NStudyCommits{} commits from \NCommonRepos{} repositories to answer this question at the commit level.

Agent commits differ from Agent-era Human commits mainly in the length of their messages, and Human commits themselves grew longer across eras (\rqa{}).
After signatures are masked, detectors rank Agent commits above Human commits with an AUC of up to \AUCTFIDF{}, but most of this signal comes from message verbosity and change size, and even the best detector reaches a precision of only \PrecTFIDF{} at the observed Agent share (\rqb{}).
Without fine-tuning, \texttt{gpt-6-sol} reaches AUCs similar to XGBoost and TF-IDF on the shared sample.
Detection transfers poorly to agents absent from training, with the vocabulary signal falling to an AUC of \AUCUnseenCCTFIDF{} (\rqc{}), and it is unstable over time, partly because the form in which agent commits reach the default branch changes (\rqd{}).

These results show that signature-free detection can prioritize commits for review but cannot replace recorded attribution.
Studies that rely on signature-based labels should treat them as a lower bound on agent activity, report precision at the observed prevalence, and check that masking also removes merge-workflow artifacts.
In future work, we plan to validate the Human labels by manually inspecting high-scoring Human commits, to evaluate detectors that learn stronger representations of code changes, and to study how agent workflows shape the commits that reach the default branch.


\section*{Data Availability}
Our replication package~\cite{artifact_agent_or_human} contains the scripts we used to collect the data and run every analysis.
It includes the identifiers, labels, and \NFeatures{} features of all \NStudyCommits{} commits, with and without masking, along with the detectors' out-of-fold predictions and the LLM prompts and raw responses.
A single script regenerates the result tables and figures from these files.



\balance
\bibliographystyle{ACM-Reference-Format}
\bibliography{references}

\end{document}